\documentclass[nofootinbib,superscriptaddress,twocolumn,showpacs,preprintnumbers,amsmath,amssymb]{revtex4-1}

\usepackage{url}
\usepackage{graphicx}
\usepackage[update,prepend]{epstopdf}
\usepackage{epsfig}
\usepackage{grffile}
\usepackage{slashed}
\usepackage{bbm}
\usepackage{footmisc}
\usepackage{multirow}

\usepackage{hyperref}
\usepackage{amssymb}
\usepackage{amsmath}
\usepackage{color}
\usepackage{xspace}
\usepackage{caption}
\usepackage{subcaption}
\usepackage{natbib}

\usepackage{microtype}
\usepackage{siunitx}
\usepackage{booktabs}
\usepackage{cleveref}

\usepackage{datetime}
\usepackage{color}
\newcommand{\pt}{p_{\text{T}}{}}
\newcommand{\ow}{\mathcal{O}_W}
\newcommand{\owval}{\left(D_\mu \Phi \right)^\dag \hat{W}^{\mu\nu}
  \left(D_\nu \Phi \right)}

\newcommand{\etj}{\sum_{\text{jets}}\text{E}_{\text{T},i}}
\newcommand{\etew}{\sum_{W,Z/H}\text{E}_{\text{T},i}}

\newcommand{\WH}{\ensuremath{e^{\pm}\overset{\textbf{\fontsize{0.5pt}{0.5pt}\selectfont(---)}}{\nu_{e}} H}}
\newcommand{\WPZ}{\ensuremath{e^+\nu_{e}\ \mu^+ \mu^-}}
\newcommand{\WPH}{\ensuremath{e^+\nu_{e}\ H}}

\begin{document}
\title{QCD radiation in $WH$ and $WZ$ production and anomalous
  coupling measurements}

\preprint{
FTUV-14-3008\;\; IFIC/14-57\;\;
KA-TP-25-2014\;\; LPN14-109\;\;
SFB/CPP-14-66}

\author{Francisco~Campanario} \email{francisco.campanario@ific.uv.es}
\affiliation{Theory Division, IFIC, University of Valencia-CSIC,
  E-46980 Paterna, Valencia, Spain.}  \author{Robin~Roth}
\email{robin.roth@kit.edu} \author{Dieter~Zeppenfeld}
\email{dieter.zeppenfeld@kit.edu} \affiliation{Institute for
  Theoretical Physics, KIT, 76128 Karlsruhe, Germany.}

\begin{abstract}
We study QCD radiation for the $WH$ and $WZ$
production processes at the LHC\@. 
We identify the regions sensitive to anomalous couplings, by considering jet
observables, computed at NLO QCD %
with the use of the Monte Carlo program {\texttt{VBFNLO}}.
Based on these observations, we propose the use of a dynamical jet veto. 
The dynamical jet veto avoids the problem of large logarithms depending
on the veto scale, hence, providing more reliable predictions and simultaneously
increasing the sensitivity to anomalous coupling searches, especially in the
$WZ$ production process.
\end{abstract}


\maketitle

\section{Introduction}

Higgs production in association with a $W$ boson is one of the
main Higgs boson production mechanisms at the LHC\@.  The LHC
experiments did not yet observe the Higgs boson in this channel, but measurements are
compatible with the Standard Model~(SM)
prediction~\cite{Chatrchyan:2013zna,TheATLAScollaboration:2013lia}.

$VH$ production is the best channel to measure the Higgs decay to  $b \bar{b}$
at the LHC since the leptons from the $V$ decay can
be used for triggering and to reduce the backgrounds.  Additionally, it
allows the study of the $VVH$ vertex and possible modifications to it by new
physics entering via anomalous couplings (AC). In this article, we will focus on $WH$ production.

From the theoretical point of view, $WH$ production has been extensively studied
in the literature and results at the next-next-to-leading order~(NNLO)
in QCD have been provided in Ref.~\cite{Brein:2003wg} at the total cross section
level and in Ref.~\cite{Ferrera:2011bk} for differential distributions.
AC effects are also a subject of interest~\cite{Ellis:2014dva-prefix}.

Due to the large gluon luminosity, 
the fraction of $WH$ events with additional jets is
large. 
Results for $WHj$ production at NLO are thus necessary, when one looks at one
jet inclusive events as 
done by ATLAS~\cite{TheATLAScollaboration:2013lia}.
Results for this process at NLO QCD have been reported
both for $W$ on-shell production~\cite{JiJuan:2010ga}
and also including the leptonic
decays of the $W$~\cite{Luisoni:2013cuh}.

In vector boson pair production processes, it is known that additional jet
radiation reduces the sensitivity to AC measurements,
results that have been confirmed at NLO in
Refs.~\cite{Campanario:2010xn,Campanario:2010hv}. 
To reduce this effect and the sensitivity to higher QCD corrections,
the traditional method has been to apply a jet veto above a 
fixed $\pt$~\cite{Baur:1993ir}, which comes
with a naive reduction of scale dependence at the total cross section
level. 
A closer look at the scale uncertainties in differential
distributions reveals that exclusive samples inherit large scale
uncertainties in the tails of the distributions, which are the regions
most sensitive to
AC effects. This has also been confirmed using merged samples
for $WW(WZ)$ and $WWj(WZj)$ using the \texttt{LOOPSIM}
method~\cite{Campanario:2012fk,Campanario:2013wta}. Thus, more
sophisticated strategies  in current Monte-Carlo driven analysis are needed to
gain theoretical control.

In this paper, we study the jet radiation patterns at NLO QCD in $WH$ and $WZ$
production.
We will show that they have distinctive signatures  and we will present
a possible strategy to increase the sensitivity in those channels to
AC searches.
The constructed jet observables are shown at NLO QCD\@.
To accomplish this, we have computed $WH(j)$ production at NLO QCD, including
Higgs and leptonic $W$ decays, and with the possibility to switch on AC effects.
These processes are available in 
\texttt{VBFNLO}~\cite{Arnold:2008rz,*Arnold:2011wj,*Baglio:2014uba}, a
parton level Monte Carlo program which allows the definition of general
acceptance cuts and distributions.

The paper is organized as follows. In Section~\ref{sec:calc_details},
the details of our calculation are given. Numerical results, including new
strategies to enhance the sensitivity to AC searches will be given in
Section~\ref{sec:results}. Finally, in Section~\ref{sec:con}, we present our
conclusions. 


\section{Calculational setup}
\label{sec:calc_details}
The $WZ/WZj$ samples at NLO QCD are obtained from
Refs.~\cite{Bellm:Thesis:2012,Campanario:2010hp} available in 
the \texttt{VBFNLO}
package~\cite{Arnold:2008rz,Arnold:2011wj,Baglio:2014uba}.
The NLO QCD corrections to $WZ$ production were first calculated in Ref.~\cite{Ohnemus:1991gb}.
To compute the $WH(j)$ production processes at NLO QCD, we 
simplified the calculation for
$l \nu_l \gamma \gamma j$~\cite{Campanario:2011ud} production (from now on
called $W \gamma \gamma j$ for simplicity)
as explained below.
For more details on the implementation and the checks performed, we refer the
reader to Ref.~\cite{Roth:Thesis:2013}. There, also, comparisons to earlier
calculations of NLO QCD corrections to $WHj$ production~\cite{JiJuan:2010ga,Luisoni:2013cuh} are discussed.
In the following, we sketch some details of our approach 
to make this work self-contained.

To compute the LO, virtual and real  corrections, we use the effective
 current approach and the spinor-helicity amplitude
method~\cite{Hagiwara:1988pp,Campanario:2011cs} factorizing the leptonic
tensor containing the EW information of the system from the QCD amplitude.
This allows us to obtain the code for the $WHj$ process  from the $W\gamma
\gamma j$ code.   For the $l \nu_l \gamma \gamma j$ process, first,  the generic
amplitudes $ W  \gamma \gamma j$, $ \hat{W} \gamma j$  and $\tilde{W} j$ are
created. Then, the leptonic decays $W \to l^+ \nu $, $\hat{W} \to l^+ \nu \gamma
$ and $\tilde{W} \to l^+ \nu \gamma \gamma $ are included via effective currents,
incorporating, in this way, all off-shell effects and spin correlations of the process.

For obtaining the $WHj$ amplitude, we select the generic $\tilde{W} j$
amplitude from the $W \gamma \gamma j$ process, and use the appropriate leptonic
current, i.e.,  $\tilde{W} \to l^+ \nu H $. Decays of the Higgs boson factorize
and can be included via branching ratios for the
$H \rightarrow ff$ channels and via effective currents for the $H \rightarrow
4l$ ones. In this paper, an on-shell Higgs is assumed since
off-shell effects contribute at the level of $\num{e-3}$ and thus are negligible.
For the $WH$ process, we proceed in
a similar way starting from $W\gamma$ production~\cite{Bellm:Thesis:2012}.

These changes are global in our code and have been cross-checked by comparing
the LO and real emission corrections against
Sherpa~\cite{Gleisberg:2008ta,Gleisberg:2008fv}. Agreement at the per mille level was found for
integrated cross sections. %

Quark mixing effects as well as the possibility to choose between
the 4-flavour and 5-flavour scheme are available
in the $WH(j)$ production process, but not for the $WZ(j)$ channel.
Thus, for the sake of comparison, we use a unitary
CKM matrix and work in the 5-flavour scheme. Subprocesses with external top
quarks are excluded since they are considered to be a different process, but
virtual top-loop contributions are included in our calculation. They
contribute at the few percent level at most.


Using the Effective Field Theory formalism, the electroweak vertices such
as the ones appearing in $WZ(j)$ and $WH(j)$ production can
be extended to account for beyond the SM physics. These effects are
constructed as additional  
terms in the Lagrangian with dimensionful couplings,
\begin{equation}
	\mathcal{L}=\mathcal{L}_{\text{SM}} + \sum_i \frac{f_i}{\Lambda^2} \mathcal{O}_i\,.
	\label{eq:lagrangian-ac}
\end{equation}
We use the basis presented in Refs.~\cite{Buchmuller:1985jz,Hagiwara:1993ck} to
parameterize the AC\@. In our code, the anomalous trilinear couplings are
included via purpose-built
Helas routines which are again incorporated via effective currents, e.g.\ for
$WH(j)$ production, we replaced  $\tilde{W} $, by $\tilde{W} = \tilde{W}^{SM}
+\tilde{W}^{AC} $, with  $\tilde{W}^{SM} \to l^+ \nu H $ and $\tilde{W}^{AC}
\to l^+ \nu H $, the latter via AC contributions coming from dimension 6 operators.
The ones needed for the $WZ(j)$ production process were included in
\texttt{VBFNLO} in a
dedicated study in Ref.~\cite{Campanario:2010xn}.
The specific ones for $WH(j)$ production were first included in Ref.~\cite{Hankele:2006ma}.

While all relevant operators are implemented, we will focus the discussion on
the operator 
\begin{align}
	\ow &= \owval,
	\label{eq:ow}
\end{align}
which does not only induce anomalous $VVH$ couplings, but also introduces modifications in
$WWV$ vertices and, thus, is severely constrained by LEP data already.
The global fit of Ref.~\cite{Corbett:2012ja} bounds the coupling in the 
range  $f_W/\Lambda^2 \in [-5.6, 9.6]\ \si{TeV^{-2}}$,
which is slightly more restrictive than the fit presented in
Ref.~\cite{Masso:2012eq}. 

In general, the AC contribution is most pronounced at large $WH$ 
invariant mass.
To measure AC effects,  all contributing operators have to be considered,
but we will focus and use $\ow$ as a typical
representative in the following. 
%
Note that there are remarkable differences in the coupling
structure induced by $\ow$ in the $WWH$ and $WWZ$ vertex.

To preserve tree level unitarity, we use a dipole form factor of the type
\begin{equation}
	F=\left( 1 + \frac{s}{\Lambda_{\rm{FF}}^2} \right)^{-p},
	\label{eq:ff}
\end{equation}
with $p=1$ and $\Lambda_{\rm{FF}}=\SI{2}{TeV}$, where $\sqrt{s}$ denotes the
$WH$ or $WZ$ invariant mass. The value for the form factor scale
$\Lambda_{\rm{FF}}$ is derived from requiring that unitarity is preserved in 
$VV \rightarrow  VV$ scattering using the form factor tool available on the
\texttt{VBFNLO} website~\cite{VBFNLOFF:2014}.

\begin{table*}[ht!] 
	\setlength{\tabcolsep}{2ex}
	\centering
	\caption{Cross sections (in \si{fb}) for various jet multiplicities at LO and NLO \\
        for $\WPZ (j)$ final states, for
        inclusive and boosted cuts as defined in
				Eqs.~\ref{eq:basiccuts}~and~\ref{eq:boosted}.\\
				The relative statistical error is less than \num{3e-3}.
	}
	\begin{tabular}{lllllllll}
		\toprule
		& \multicolumn{4}{c}{inclusive} & 		\multicolumn{4}{c}{boosted} \\
		\cmidrule(r){2-5} \cmidrule(r){6-9}
		& \multicolumn{2}{c}{$W^+Z$} & 		\multicolumn{2}{c}{$W^+Zj$} & \multicolumn{2}{c}{$W^+Z$} &
		\multicolumn{2}{c}{$W^+Zj$}\\
		\cmidrule(r){2-3} \cmidrule(r){4-5} \cmidrule(r){6-7} \cmidrule(r){8-9}
		$\text{N}_{\text{jets}}$ 
		  & LO      & NLO     & LO      & NLO     & LO     & NLO     & LO      & NLO \\
			\midrule
			0 & 14.00 & 16.74 &       &       & 0.492 & 0.397  &         & \\
		1 &         & 11.28 & 11.31 & 8.391 &       & 1.242 & 1.248 & 0.554 \\
		2 &         &       &       & 6.223 &       &         &         & 1.094 \\
		\bottomrule
	\end{tabular}
	\label{tab:njetxsec}
\end{table*}
\begin{table*} 
	\setlength{\tabcolsep}{2ex}
	\centering
	\caption{Cross sections (in \si{fb}) for various jet multiplicities at
          LO and NLO \\ 
        for $\WPH (j)$ final states, for
        inclusive and boosted cuts as defined in
		Eqs.~\ref{eq:basiccuts}~and~\ref{eq:boosted}.\\
	The relative statistical error is less than \num{3e-3}.}
	\begin{tabular}{lllllllll}
		\toprule
		& \multicolumn{4}{c}{inclusive} & 		\multicolumn{4}{c}{boosted} \\
		\cmidrule(r){2-5} \cmidrule(r){6-9}
		& \multicolumn{2}{c}{$W^+H$} & 		\multicolumn{2}{c}{$W^+Hj$} & \multicolumn{2}{c}{$W^+H$} &
		\multicolumn{2}{c}{$W^+Hj$}\\
		\cmidrule(r){2-3} \cmidrule(r){4-5} \cmidrule(r){6-7} \cmidrule(r){8-9}
		$\text{N}_{\text{jets}}$ 
		  & LO      & NLO     & LO      & NLO     & LO     & NLO     & LO      & NLO \\
			\midrule
			0 & 47.08 & 44.12 &       &       & 4.103 & 3.188 &       &         \\
			1 &       & 19.72 & 20.16 & 16.12 &       & 2.648 & 2.690 & 1.889 \\
			2 &       &       &       &  7.16 &       &       &       & 1.243 \\
		\bottomrule
	\end{tabular}
	\label{tab:whjnjetxsec}
\end{table*}

\section{Numerical results}
\label{sec:results}
In the following, we present results for the LHC operating at 14 TeV
center-of-mass energy for the specific final states
$\WPH (j)$ and
$\WPZ (j)$
and refer to them respectively as $W^+H(j)$ and $W^+Z(j)$ production for
simplicity.
Results for $W^-H(j)$ and $W^-Z(j)$ production are very similar.

As input parameters,
we use $M_W=\SI{80.3980}{GeV}$, $M_Z=\SI{91.1876}{GeV}$, $M_H = \SI{126.0}{GeV}$ and
$G_F=\SI{1.16637e-5}{GeV^{-2}}$ and derive the electromagnetic coupling
constant and the weak-mixing angle from tree level relations.
All the fermions are considered massless, except the top quark with
$m_t=\SI{172.4}{GeV}$.
The resonating propagators are constructed with a constant width, 
fixed at $\Gamma_W=\SI{2.098}{GeV}$,
$\Gamma_Z=\SI{2.508}{GeV}$ and
$\Gamma_H = \SI{4.277}{MeV}$.

We use $\overline{MS}$ renormalization of the strong coupling constant
$\alpha_s$ and the CTEQ CT10 NLO parton distribution functions~\cite{Lai:2010vv} with
$\alpha_s^\text{LO(NLO)}(M_Z)=\num{0.1298}(\num{0.1180})$. The 
running of $\alpha_s$ includes 5 massless flavours, decoupling
the top-quark contribution. 
%

As a central value for the factorization and renormalization scale, we choose
\begin{align}
	\mu_{0}=\frac{1}{2} \left(\sum_{\text{partons}}
	p_{T,i} + \sum_{W, Z/H}\sqrt{p_{T,i}^2+m_{i}^2}\right),
	\label{eq:define_mu0}
\end{align}
where $m_i$ denotes the reconstructed invariant mass of the corresponding
decay leptons or the on-shell boson.

The jets are clustered using the anti-$k_t$
algorithm~\cite{Cacciari:2008gp} with a cone radius of $R=0.4$. To simulate
typical detector acceptance, we impose a minimal set of inclusive
cuts
\begin{equation}
\label{eq:basiccuts}
\begin{aligned}
		\pt_l &> \SI{20}{GeV} & \pt_{j} &>\SI{30}{GeV} & \slashed p_{\text{T}} &> \SI{30}{GeV} \\
		|\eta_j| &< 4.5  & |\eta_l| &< 2.5 &  R_{l(l,j)} &> 0.4 \\
		m_{ll} &> \SI{15}{GeV} & R_{ll}  &> 0.4 ,
\end{aligned}
\end{equation}
where the $m_{ll}$ cut is applied only to the leptons with opposite sign coming from the $Z$ boson.
To simulate $VH$ experimental searches, we will also present results
for boosted events requiring, additionally,
\begin{align}
	\label{eq:boosted}
	\pt_{Z/H} &> \SI{200}{GeV},
\end{align}
where $\pt_Z$ is the reconstructed transverse momentum of the decay leptons.

The sensitivity to AC is not evenly distributed over phase space.
A large contribution to the $WZj$ cross section comes from events 
where the $\pt$ of the $Z$ boson and the leading jet balance and the $W$ is soft, or
similarly with $W$ and $Z$ exchanged. 
Those events can be considered as EW corrections to $Vj$ production.
Because the invariant mass of the electroweak system is small, they are less
sensitive to AC effects. To suppress these events and also to reduce the
impact of higher order QCD corrections, a common approach is to apply a jet
veto at fixed $\pt$. 
However, this procedure is problematic. The veto introduces terms of
the form $\alpha_s^n  \ln^{2n} (s/\pt^2_{j,\text{veto}}) $, where $s$ represents a
typical scale of the hard process. For large values of $s$, this results in a
poor control of our perturbative predictions, which may 
translate into large scale uncertainties of the observables.
Note, however, that when studying inclusive samples the uncertainties are
frequently underestimated by a naive scale variation.
Such features have been extensively discussed for jet vetoes also in the
context of NNLO calculations of Higgs-boson production,
see e.g. Refs.~\cite{Stewart:2011cf,Banfi:2012yh}. Instead of a fixed veto, of 
jets above a fixed transverse momentum, we will consider a specific 
dynamical veto in the following which avoids large logarithms by keeping 
the veto scale proportional to the hard scale of the process.
For $WV$ production, an effective
dynamical jet veto has recently also been discussed in 
Ref.~\cite{Gieseke:2014gka} in the context of
  observing large electroweak Sudakov logarithms at high transverse momenta.

In Tables~\ref{tab:njetxsec} and~\ref{tab:whjnjetxsec}, we show the integrated
cross sections
for different jet multiplicities appearing at different orders of perturbation 
theory for the inclusive and the boosted 
set of cuts.
One observes that a large fraction of the $WZ$ production cross section is
due to events with jets with $p_{Tj}>30$~GeV. 
With boosted cuts, the NLO sample of $WZj$ production has twice as many two
jet events than one jet events.  For $WHj$ production, jet radiation is also
significant. Thus, it is necessary to understand their radiation pattern and
how to enhance regions sensitive to AC.

%
%

\begin{figure*}[ht]
	\centering
	\begin{subfigure}[htp]{\columnwidth}
		\centering
		\includegraphics[width=\columnwidth]{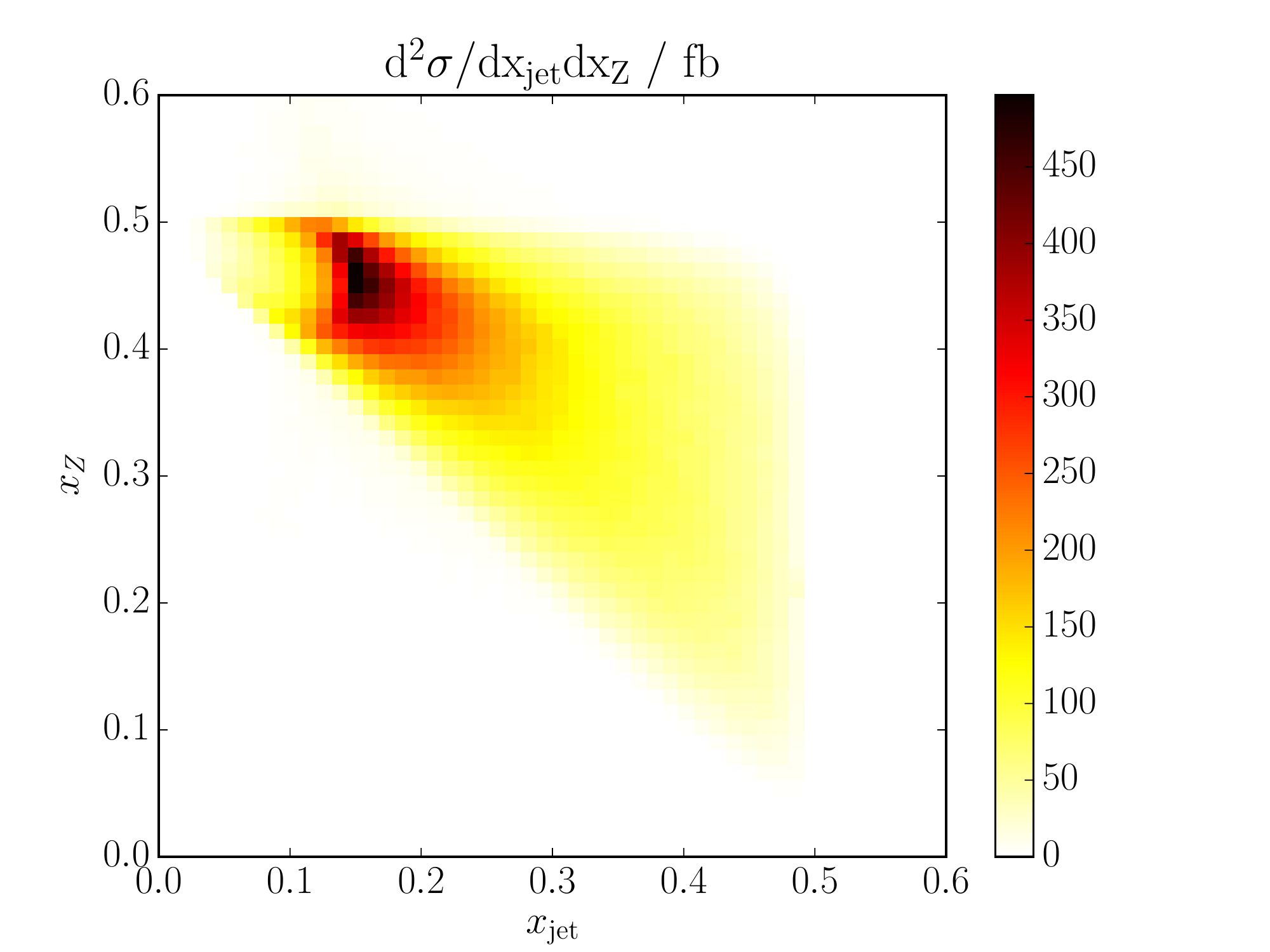}
		\caption{Inclusive $W^+Zj$}
	\end{subfigure}
	\begin{subfigure}[htp]{\columnwidth}
		\centering
		\includegraphics[width=\columnwidth]{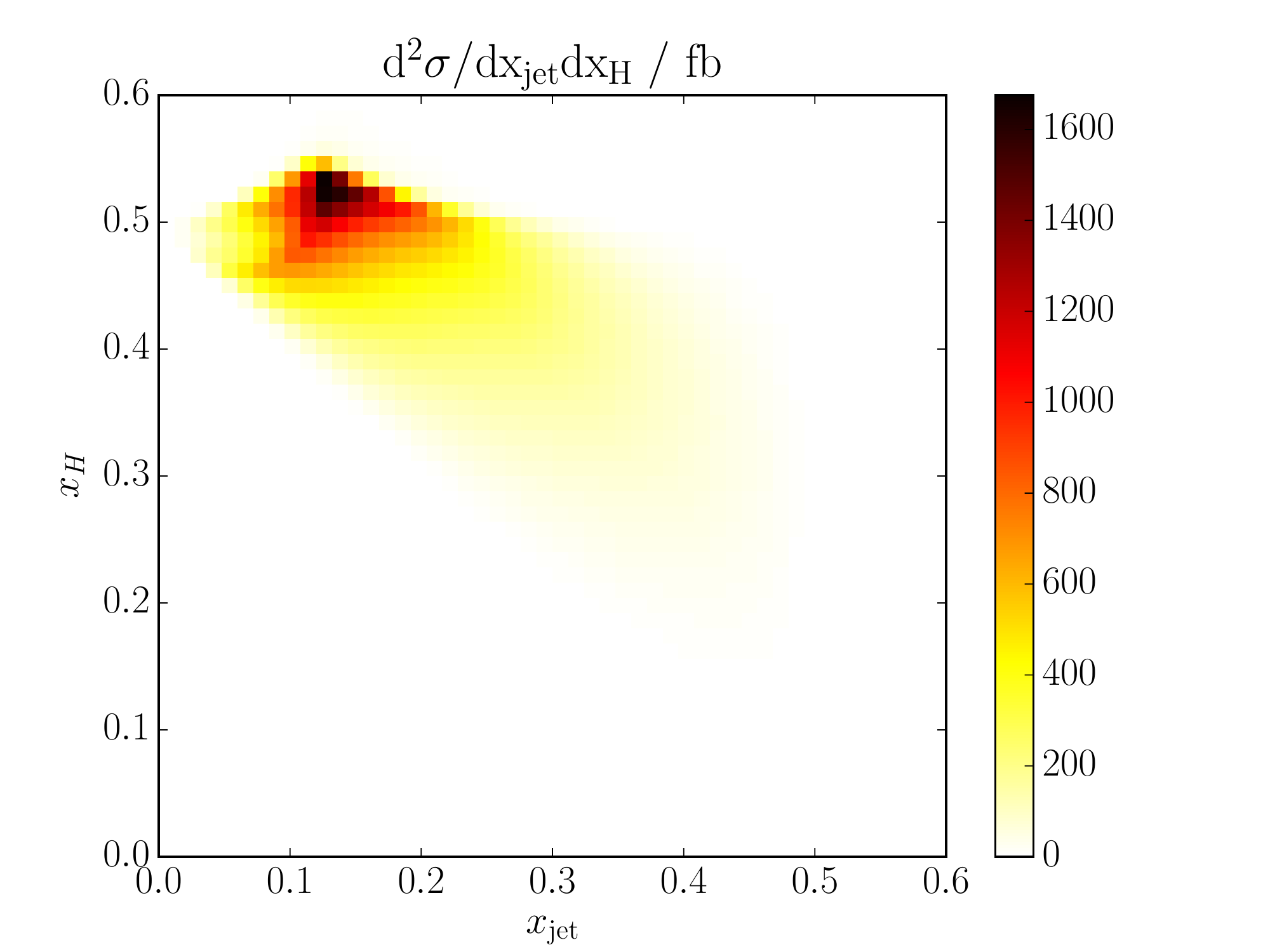}
		\caption{Inclusive $W^+Hj$}
	\end{subfigure}
	\begin{subfigure}[htp]{\columnwidth}
		\centering
		\includegraphics[width=\columnwidth]{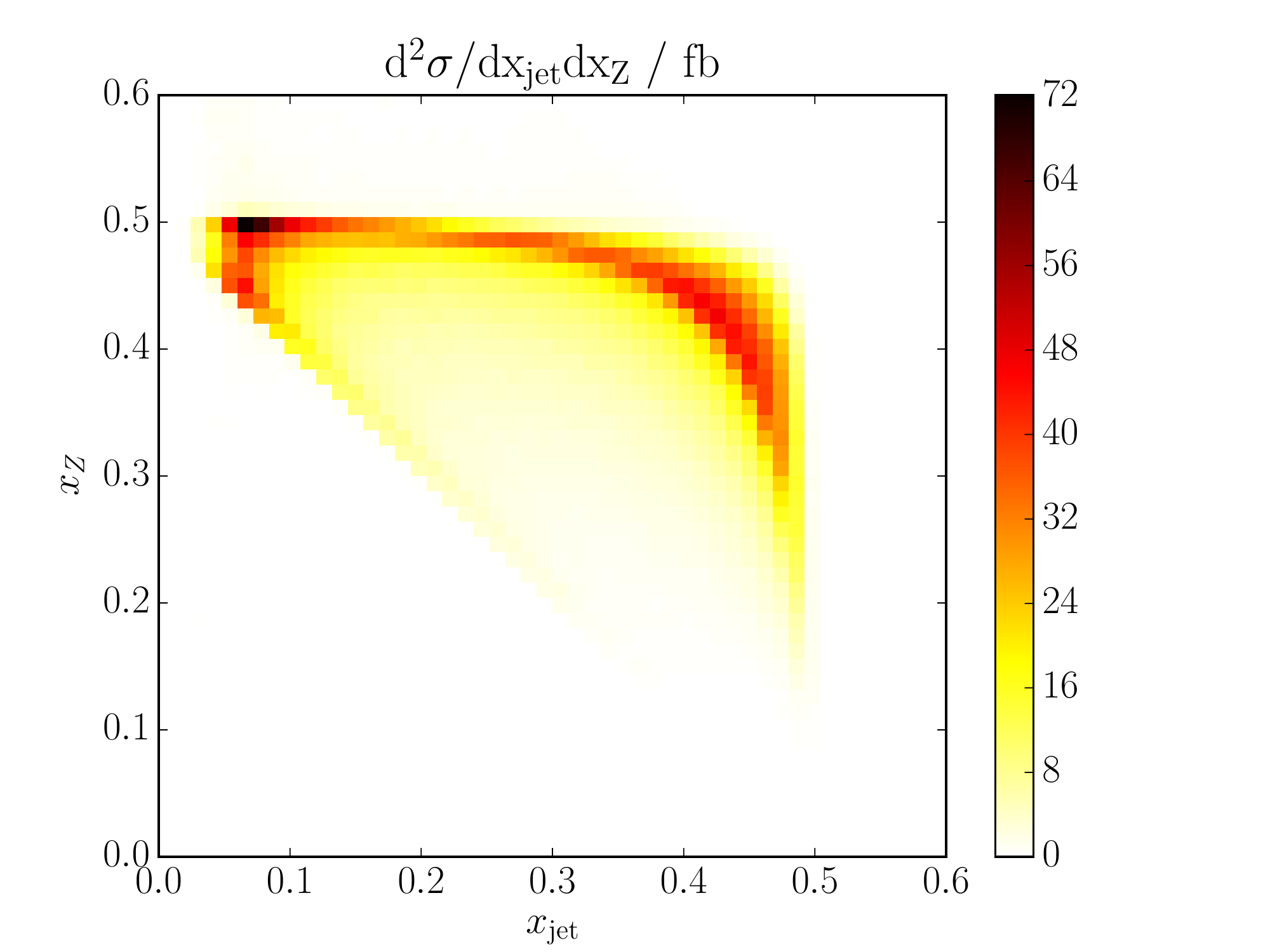}
		\caption{Boosted $W^+Zj$}
	\end{subfigure}
	\begin{subfigure}[htp]{\columnwidth}
		\centering
		\includegraphics[width=\columnwidth]{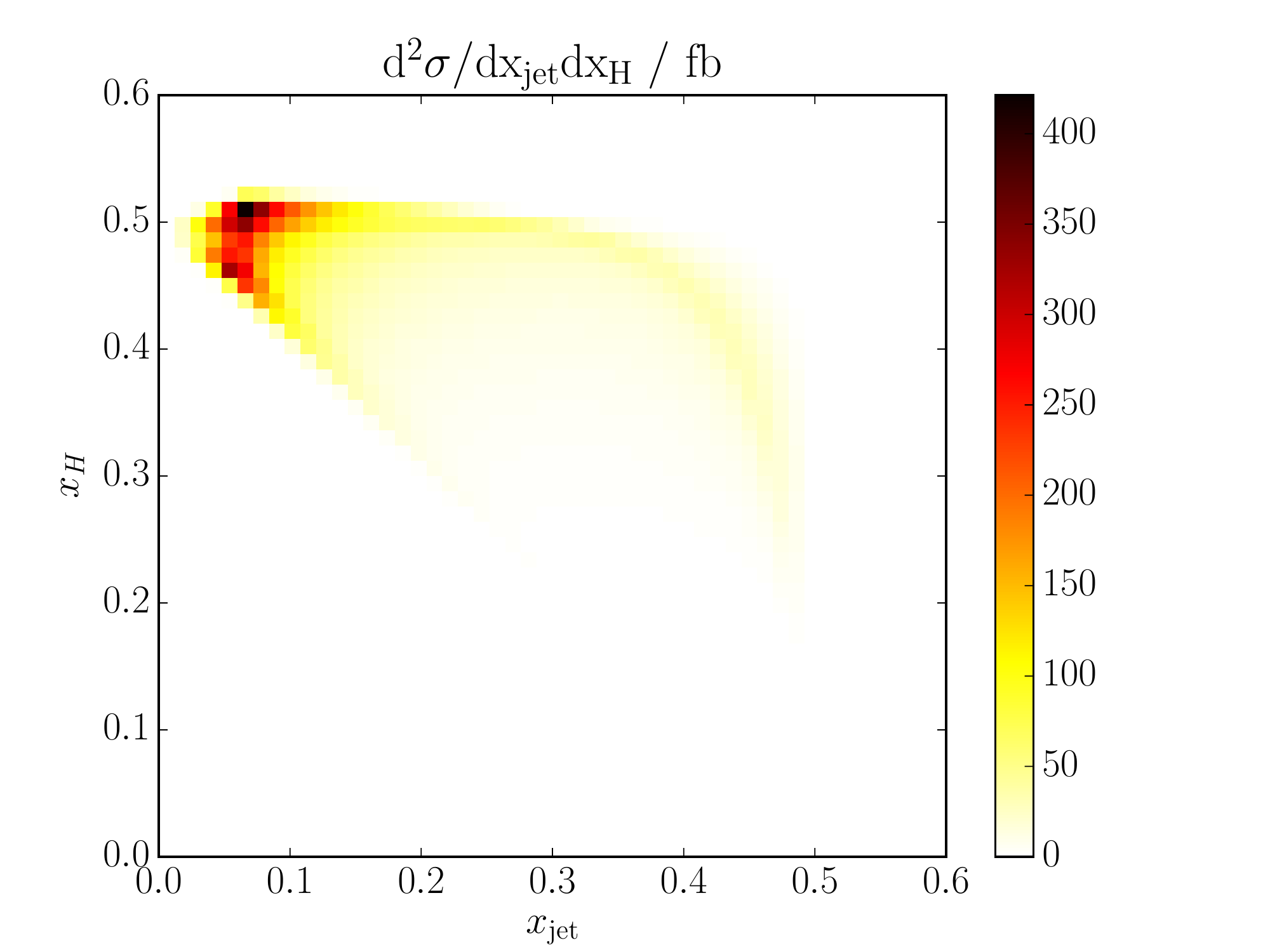}
		\caption{Boosted $W^+Hj$}
	\end{subfigure}
	\caption{LO double differential distributions for
          $\WPZ j$ ($\WPH j$) production on the left (on the right) 
          with respect to $x_{\text{jet}}$ and $x_{Z}$ ($x_{H}$). 
          Inclusive cuts are used in the upper row and boosted cuts 
					($\pt_{H(Z)} > \SI{200}{GeV}$) in the
          bottom panels.} 
\label{fig:xjLO}
\end{figure*}

In order to visualize the phase space distributions of jets and weak bosons 
and their relative hardness, it is necessary to consider their 
transverse momenta in aggregate. In a $WVj$ events at LO, transverse
momentum conservation implies 
${\bf p}_{TW}+{\bf p}_{TV}+{\bf p}_{Tj}=0$, i.e. there are four
independent transverse momentum components. Discounting an overall rotation in
the transverse plane and anticipating approximate invariance of radiation 
patterns under rescaling
at very high energies, we are left with two parameters describing the
essential features of the transverse motion and the relative importance of QCD
radiation. These can be taken as the transverse energies of two of the three 
objects, normalized to the sum for all three, i.e. we consider   
\begin{align}
	x_{\text{jet}} = \frac{\etj}{\etj+\etew}\,
	\label{eq:xjdefet}
\end{align}
and, similarly, for $V\in (W,Z,H)$ we define 
\begin{align}
	x_V = \frac{E_{TV}}{\etj+\etew}\, .
	\label{eq:xhzdefet}
\end{align}
Obviously, $x_{\text{jet}}+x_W+x_{Z/H}=1$, and in a LO calculation, where a
single massless parton forms the jet system which recoils against the other two
objects, $x_{\text{jet}}<0.5$. 

A similar definition can be constructed using transverse momentum instead of
transverse energy. However, in that case, $x_{\text{jet}}$  is
infrared sensitive and problematic in a fixed order calculation, as will be
discussed later.

\subsubsection*{$x_{\rm{jet}}$ versus  $x_{Z/H}$ Distributions}
We can use these observables to draw a Dalitz-like 2D plot of $x_{\text{jet}},
x_{Z/H}$ where phase space 
regions with soft EW bosons can easily be distinguished from regions with
soft jets. A value close to 0.5 for 
the $x_{\text{jet}(H,Z)}$ observable would indicate that the given particle
 has half of the total transverse energy 
of the system, recoiling against the rest,
while values close to zero indicate 
that the particle is soft. 


In Fig.~\ref{fig:xjLO}, we show the LO double 
differential distributions for $WZj$ and $WHj$ production with respect to 
$x_{\text{jet}}$ and $x_{Z/H}$ for inclusive (upper row) and boosted cuts
(lower panels). On the left, $WZj$ production is shown and on the right
results for the $WHj$ process, replacing $x_Z$ by the equivalent $x_{H}$ 
observable.


\begin{figure*}[htp] 
	\centering
	\begin{subfigure}[htp]{1\columnwidth}
		\centering
		\includegraphics[width=\columnwidth]{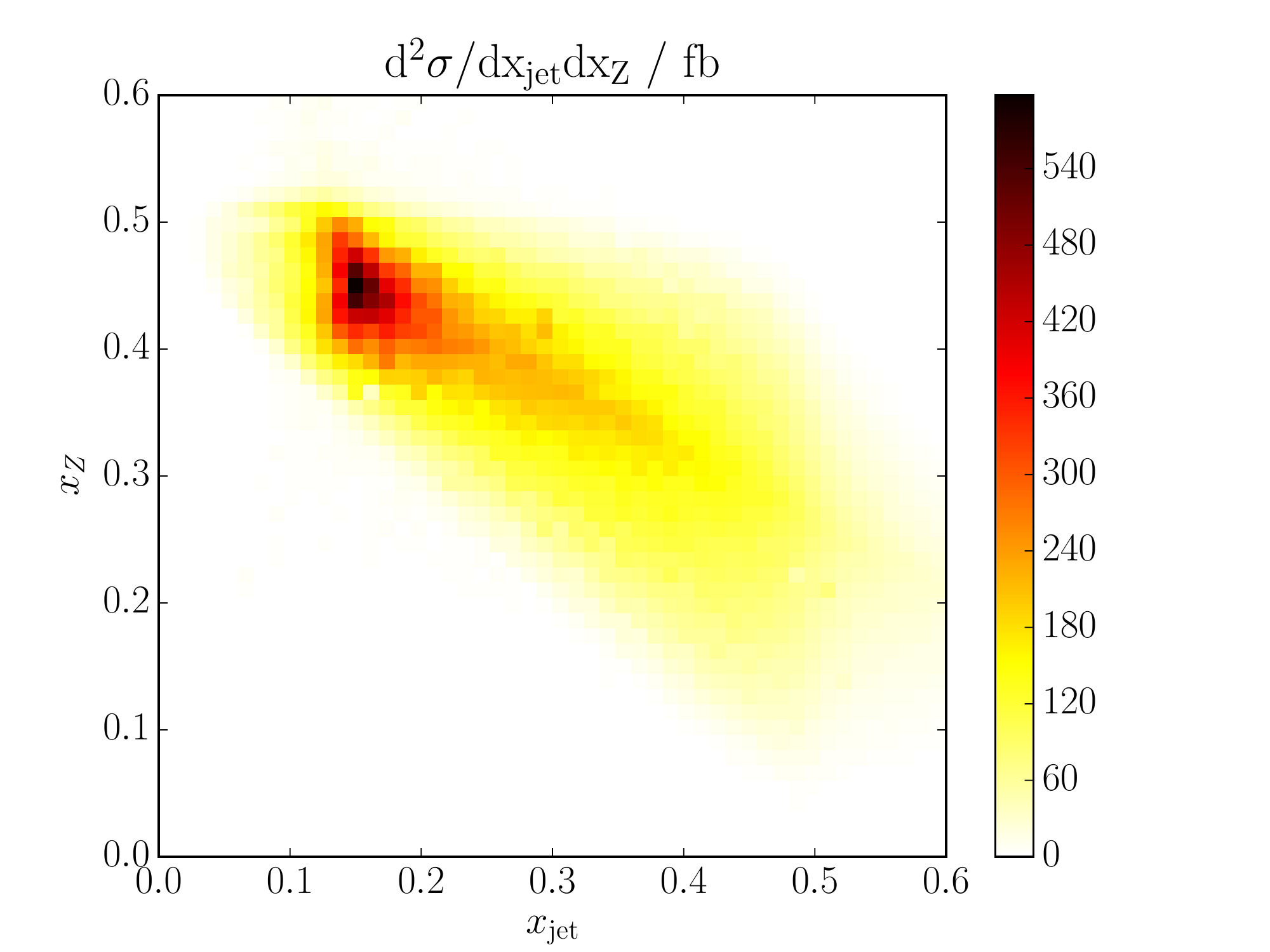}
		\caption{Inclusive $W^+Zj$}
	\end{subfigure}
	\begin{subfigure}[htp]{1\columnwidth}
		\centering
		\includegraphics[width=\columnwidth]{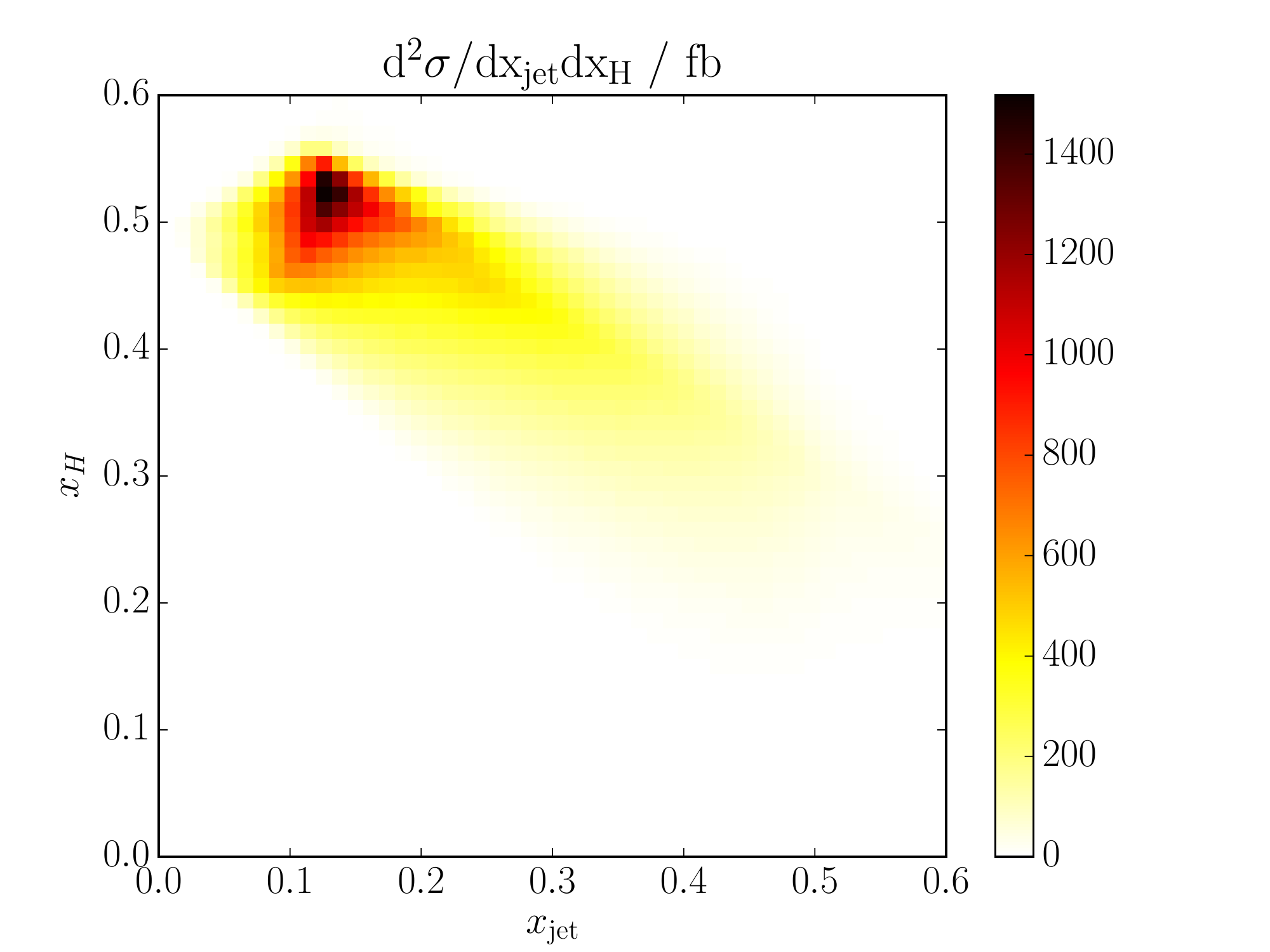}
		\caption{Inclusive $W^+Hj$}
	\end{subfigure}
	\begin{subfigure}[htp]{1\columnwidth}
		\centering
		\includegraphics[width=\columnwidth]{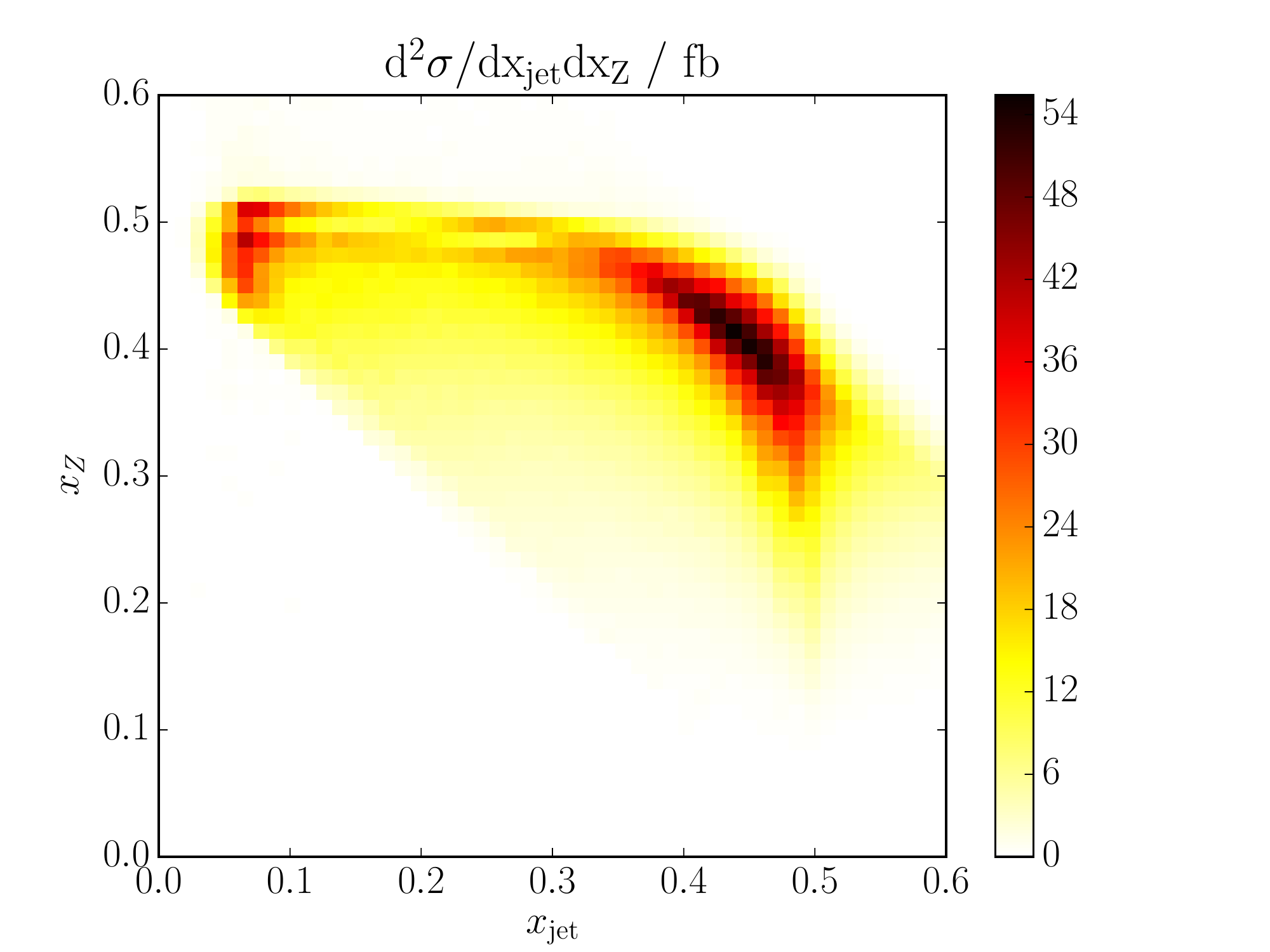}
		\caption{Boosted $W^+Zj$}
	\end{subfigure}
	\begin{subfigure}[htp]{1\columnwidth}
		\centering
		\includegraphics[width=\columnwidth]{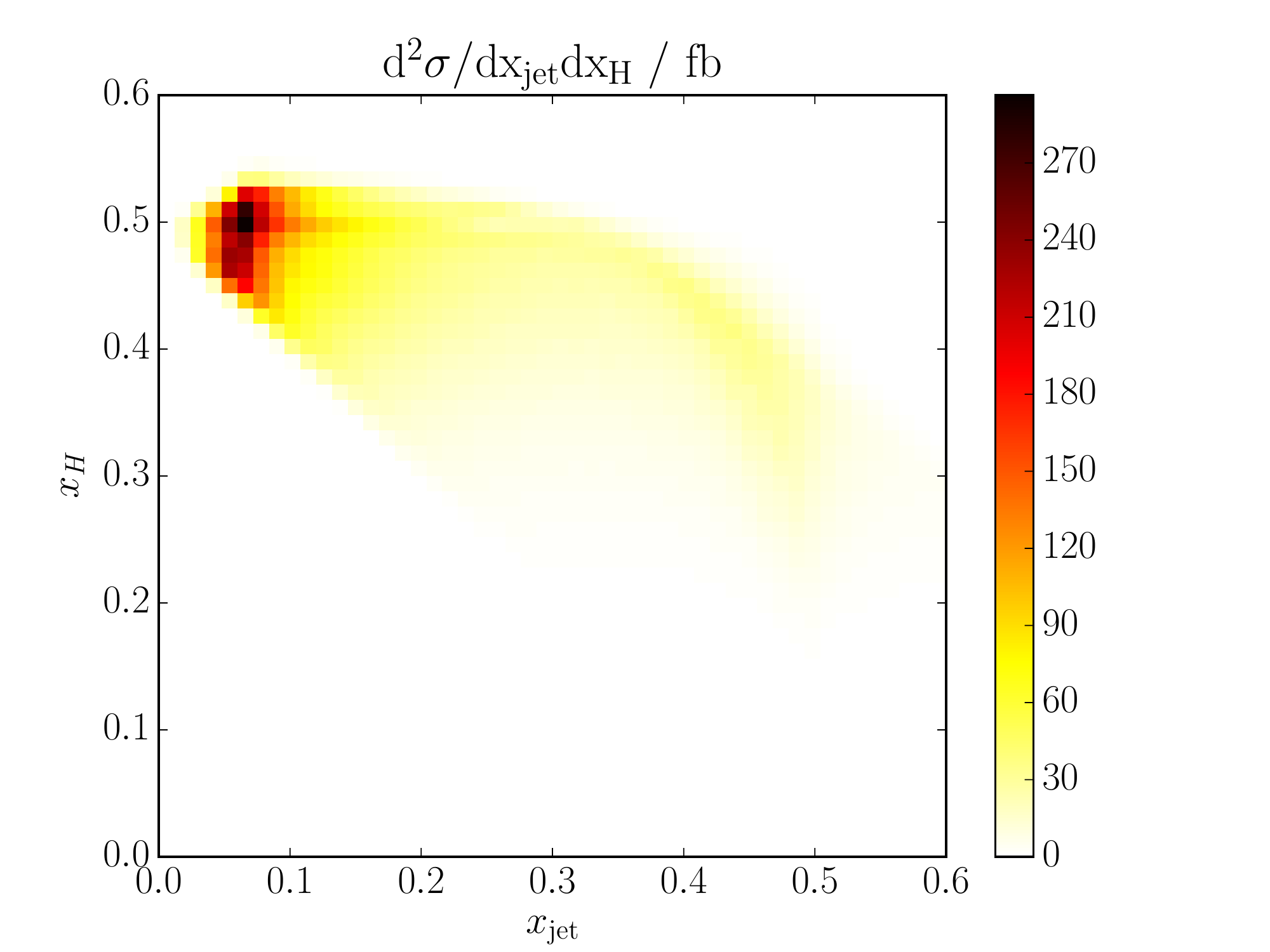}
		\caption{Boosted $W^+Hj$}
	\end{subfigure}
	\caption{Same as Fig.~\ref{fig:xjLO} but at NLO.
}
\label{fig:xjNLO}
\end{figure*}
\begin{figure*}[htp] 
	\centering
	\begin{subfigure}[htp]{\columnwidth}
		\centering
		\includegraphics[width=\columnwidth]{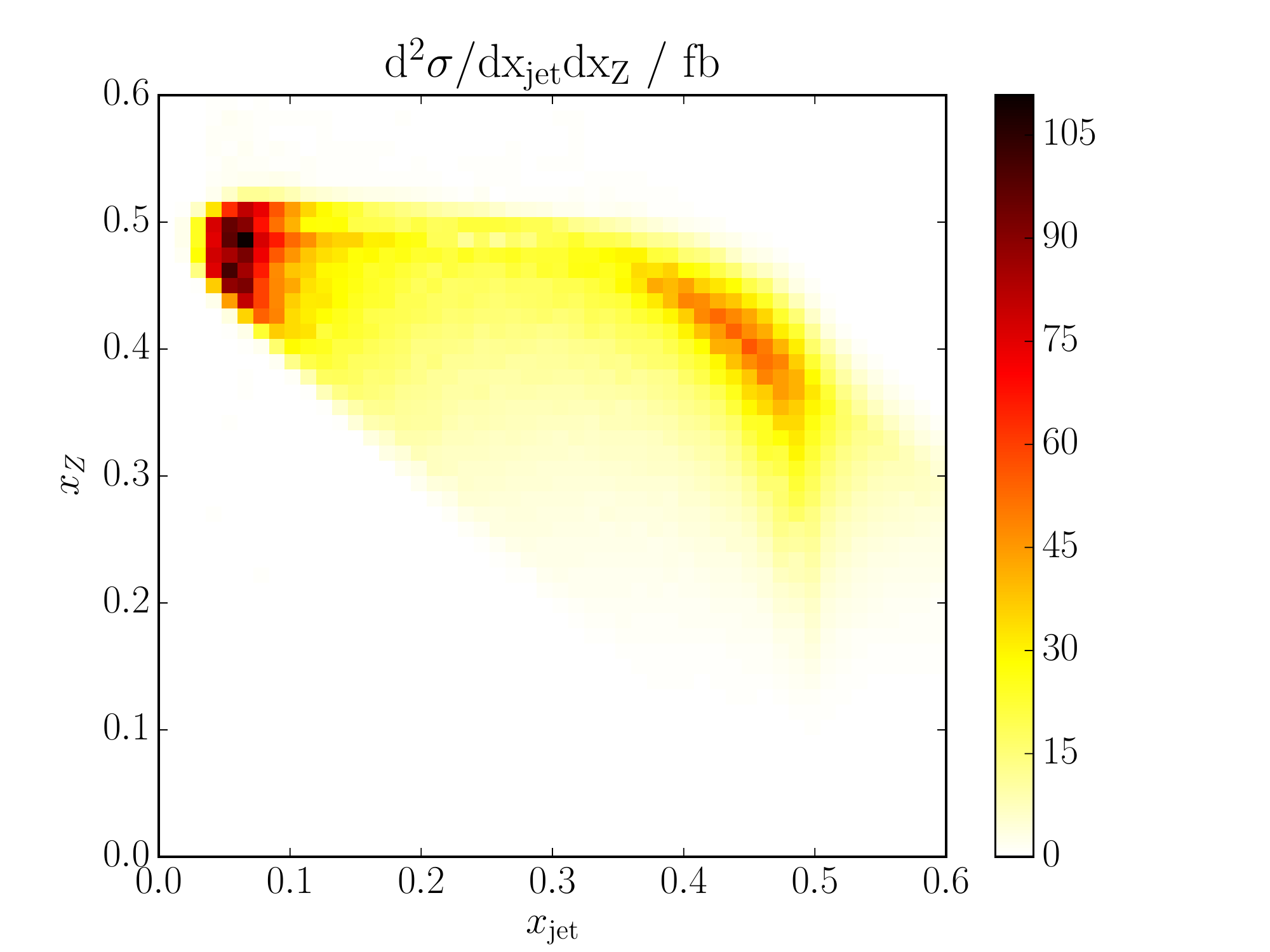}
	\end{subfigure}
		\begin{subfigure}[htp]{\columnwidth}
		\centering
		\includegraphics[width=\columnwidth]{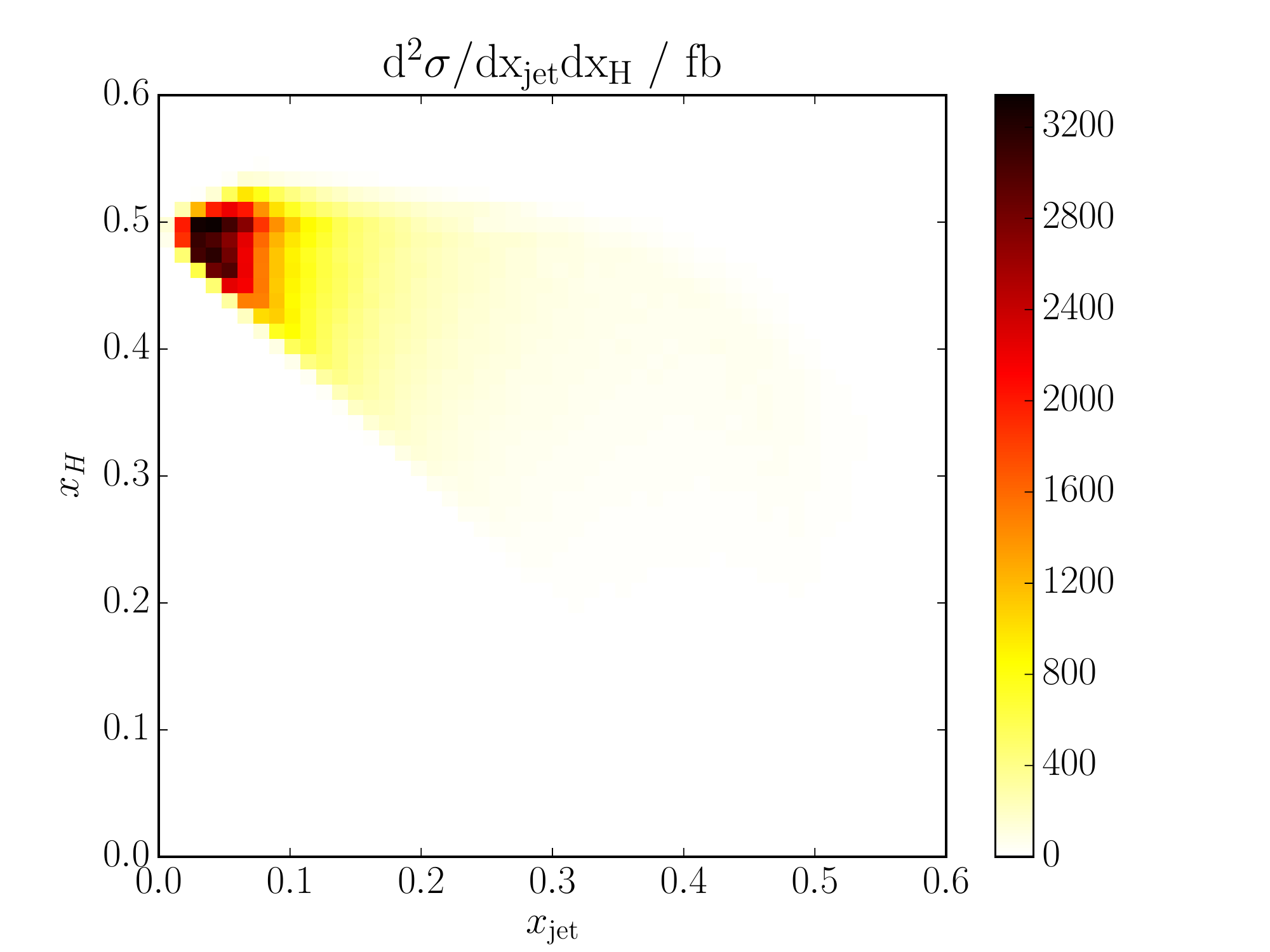}
	\end{subfigure}
	\caption{Same as Fig.~\ref{fig:xjLO} but at NLO and with anomalous
          coupling effects switched on with
          $f_W/\Lambda^2=\SI{-10}{TeV^{-2}}$.
          Results are shown for the boosted set of cuts for
	$\WPZ j$ (left panel) and $\WPH j$ production (right panel).
	}
\label{fig:xjNLOAC}
\end{figure*}
\begin{figure*}[ht!] 
	\centering
	\begin{subfigure}[htp]{\columnwidth}
		\centering
		\includegraphics[width=1.1\columnwidth]{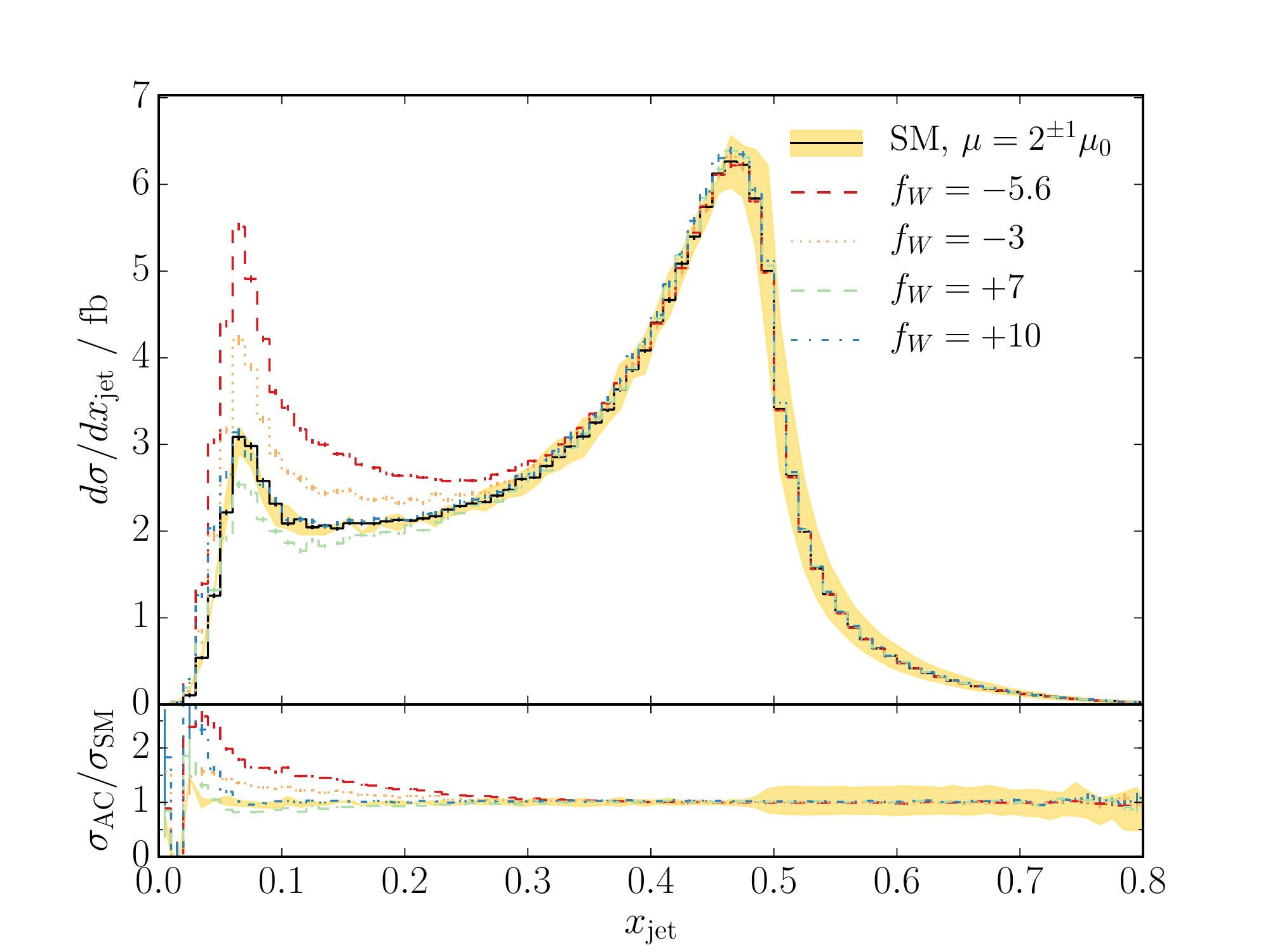}
	\end{subfigure}
	\begin{subfigure}[htp]{\columnwidth}
		\centering
		\includegraphics[width=1.1\columnwidth]{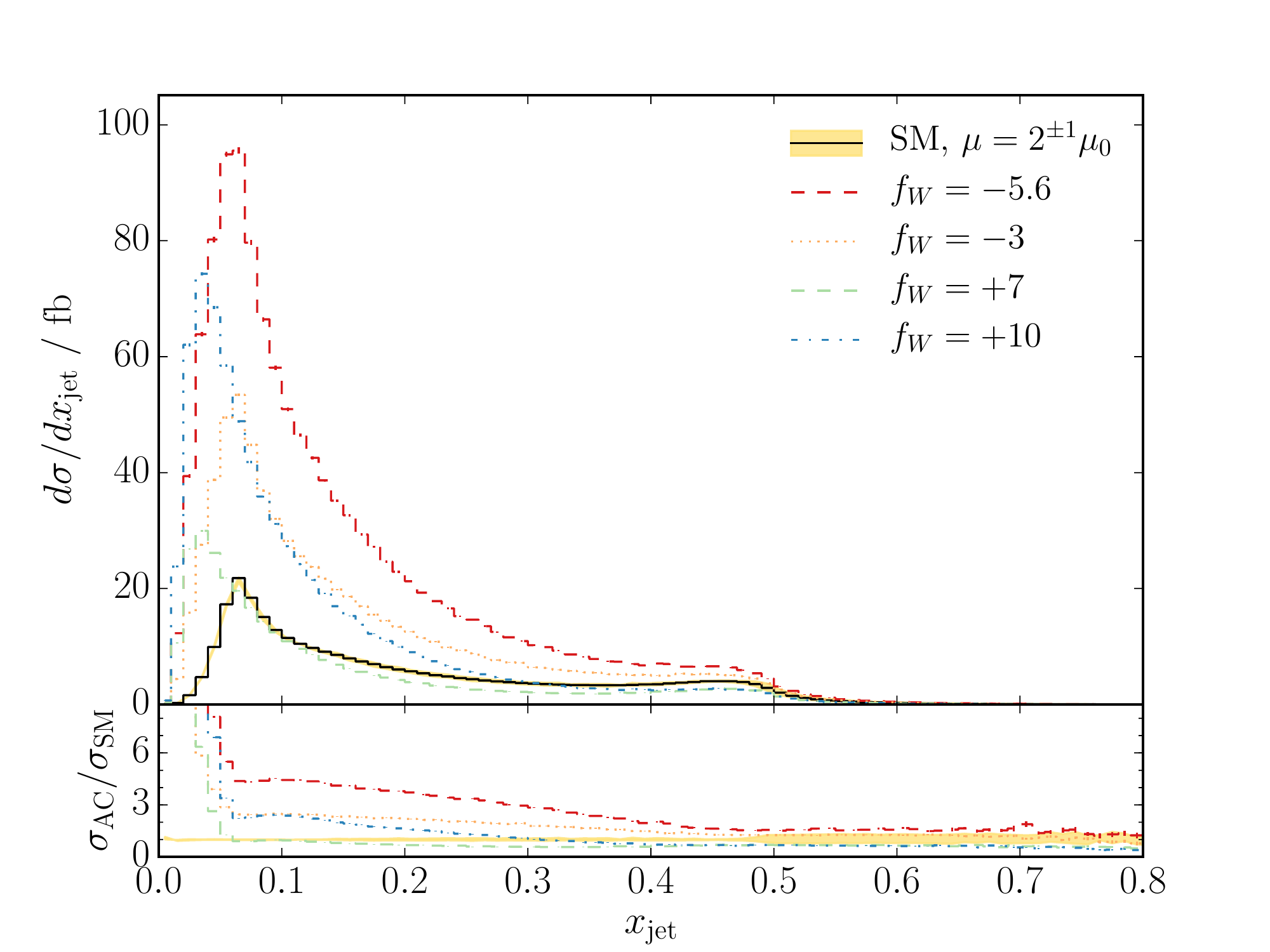}
	\end{subfigure}
	\caption{$x_{\text{jet}}$ distribution for
          $\WPZ j$ (left) and
          $\WPH j$ (right) production for different values of AC with
	boosted cuts at NLO\@.
	The yellow band corresponds to the variation of $\mu = \mu_{\rm{F}} = \mu_{\rm{R}}$ by a factor
	of 2. 
	The error bars represent the statistical Monte Carlo error.
}
	\label{fig:whj-wzj-xj-ac}
\end{figure*}

Already the inclusive sample shows that $WZj$ production allows for harder
jets, while $WHj$ production is dominated by back-to-back $WH$ pairs with only
soft jets. With the additional boosted cut, the difference is enhanced and one
can clearly observe the different radiation patterns of the $WHj$ and $WZj$
processes. %
While in $WHj$ production soft QCD radiation is preferred, in the $WZj$ case
there are two equally important phase space regions, those with soft jets at
small $x_{\text{jet}}$ and those with a soft $W$ boson at large
$x_{\text{jet}}$. The latter dilutes the sensitivity to AC of this process as
will be shown below. %
The origin of these different radiation patterns is the partial wave
decomposition of the $WH$/$WZ$ final state. $WH$ production is mostly
restricted to $J=1$, since it arises from a virtual $W$, 
while this is only a small contribution 
to $WZ$ production.

\begin{figure*}[htp]
	\centering
	\begin{subfigure}[htp]{\columnwidth}
		\centering
		\includegraphics[width=\columnwidth]{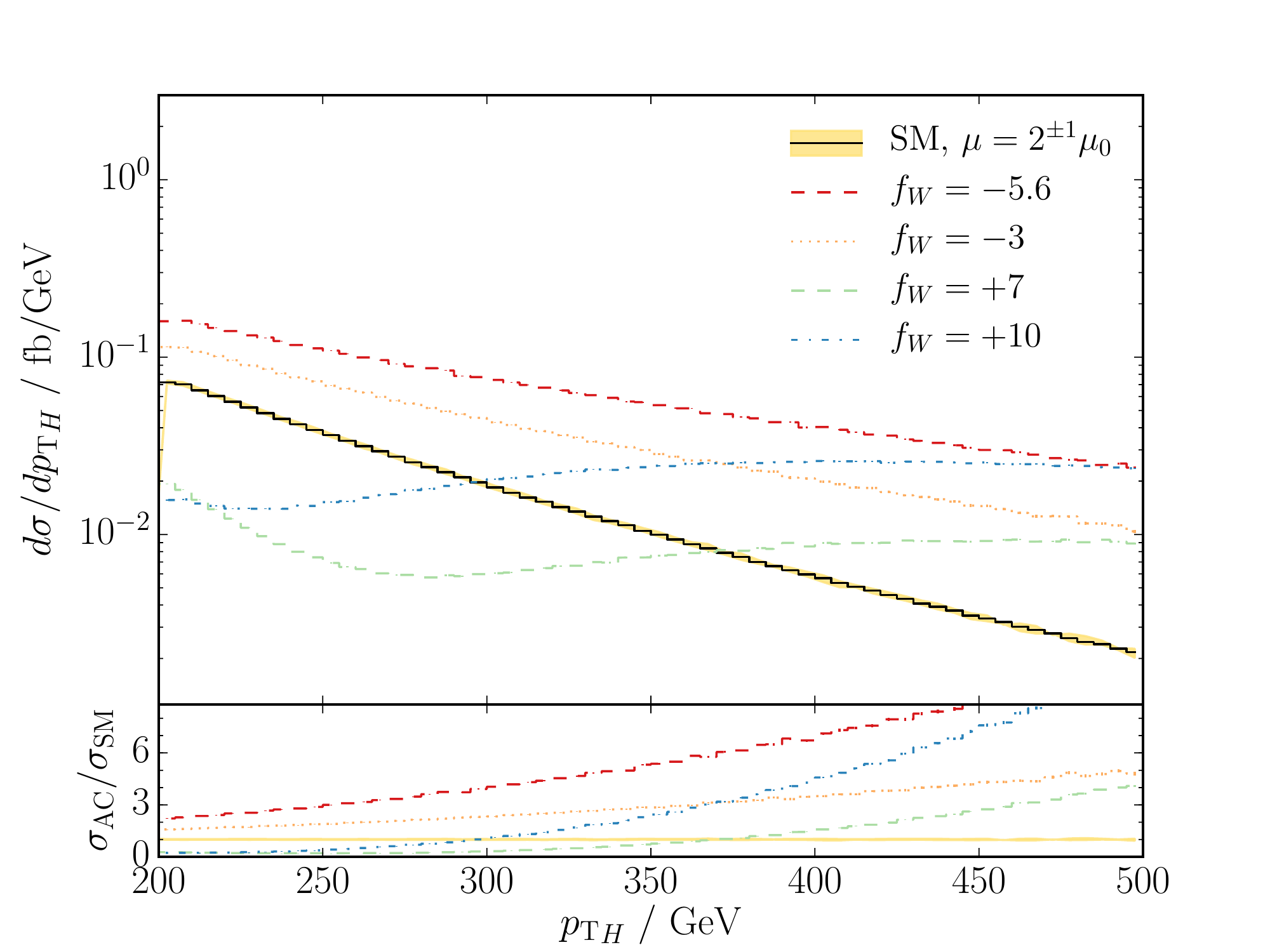}
		\caption{$\pt_H$, $W^+H$, boosted}
		\label{fig:wh-boost200-ac-ptz-nlo}
	\end{subfigure}
	\begin{subfigure}[htp]{\columnwidth}
		\centering
		\includegraphics[width=\columnwidth]{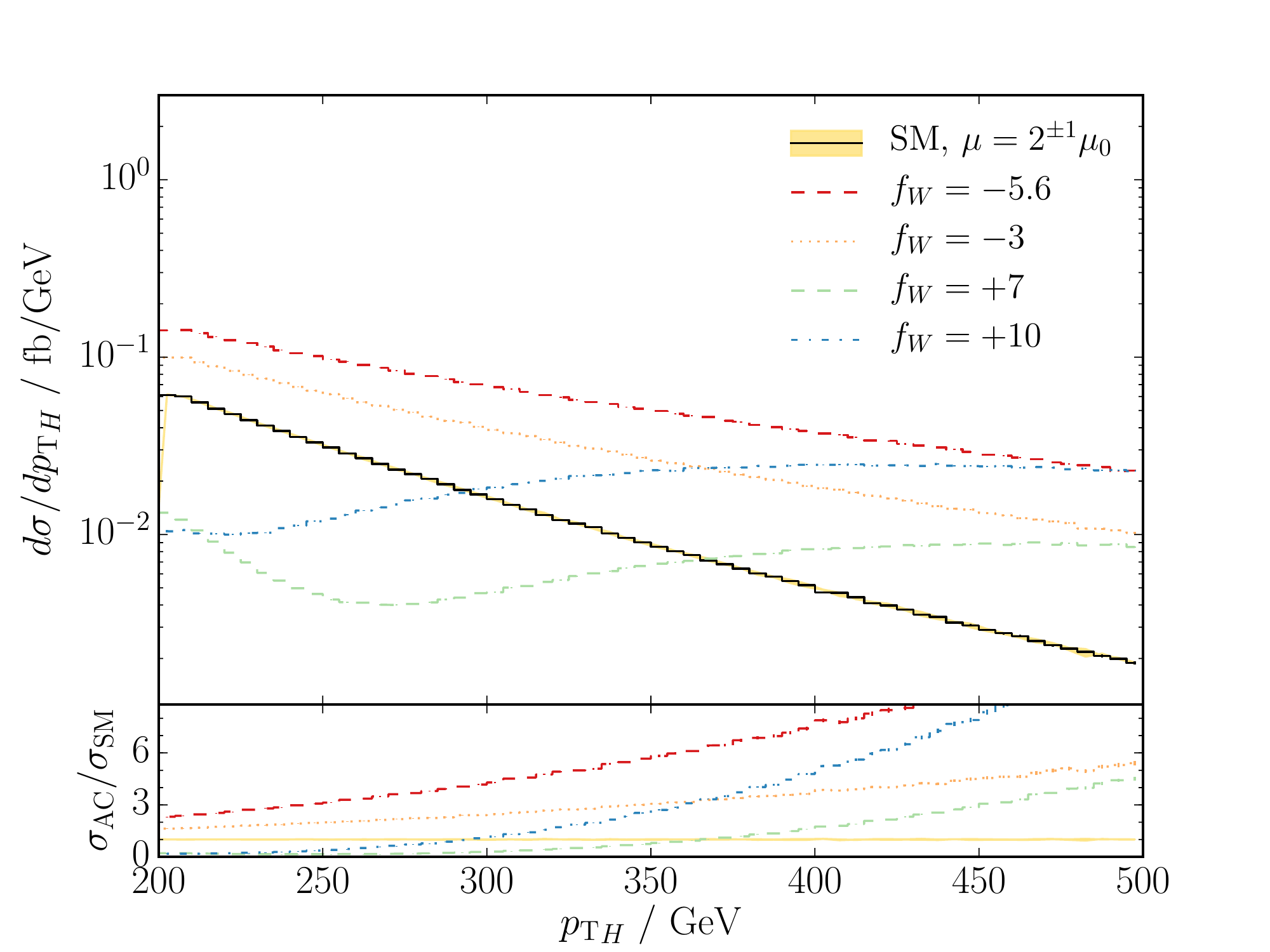}
		\caption{$\pt_H$, $W^+H$, boosted, with jet veto}
		\label{fig:wh-boost200-ac-ptz-nlo-lt02}
	\end{subfigure}
	\centering
	\begin{subfigure}[htp]{\columnwidth}
		\centering
		\includegraphics[width=\columnwidth]{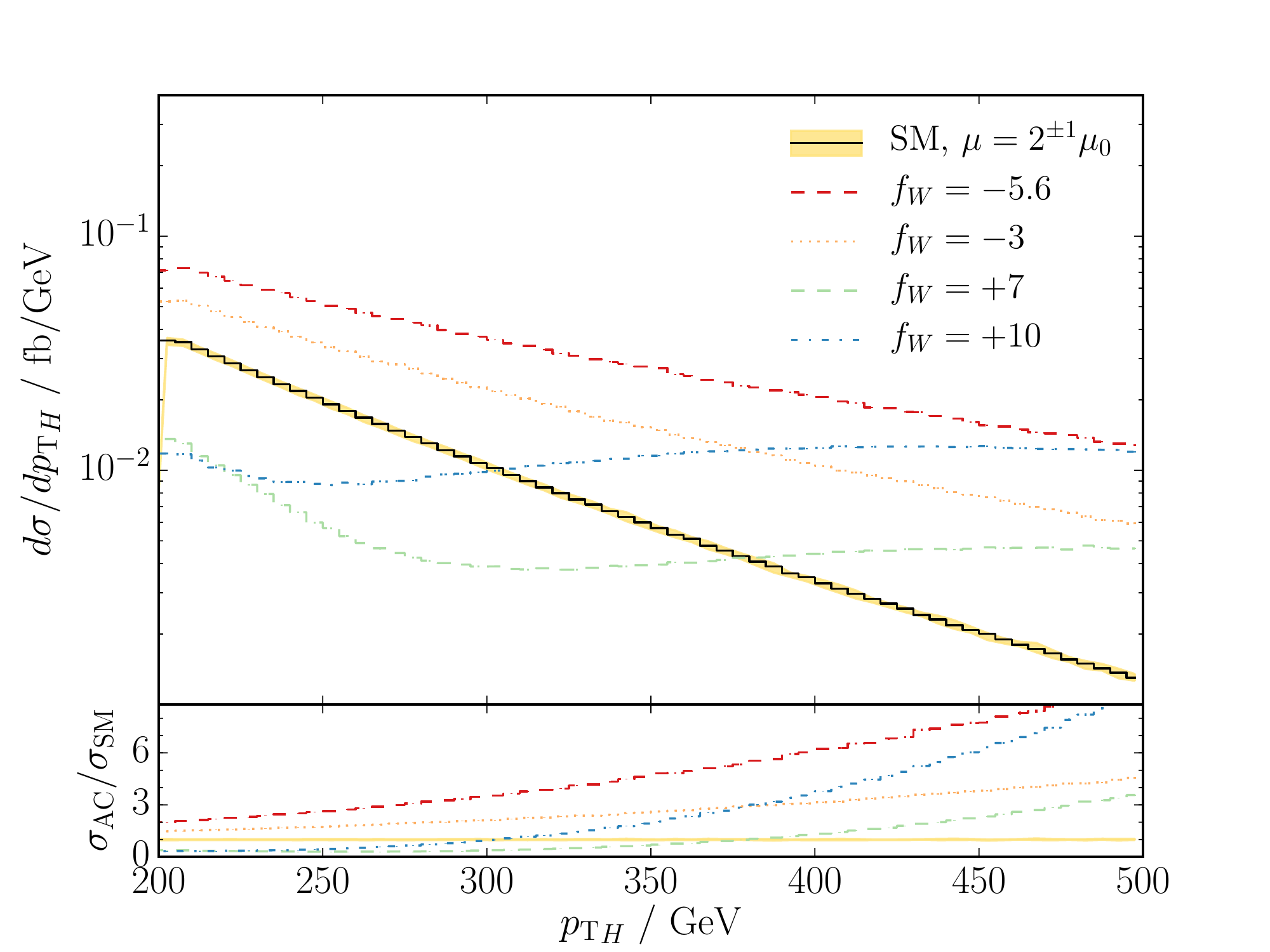}
		\caption{$\pt_H$, $W^+Hj$, boosted}
		\label{fig:whj-boost200-ac-ptz-nlo}
	\end{subfigure}
	\begin{subfigure}[htp]{\columnwidth}
		\centering
		\includegraphics[width=\columnwidth]{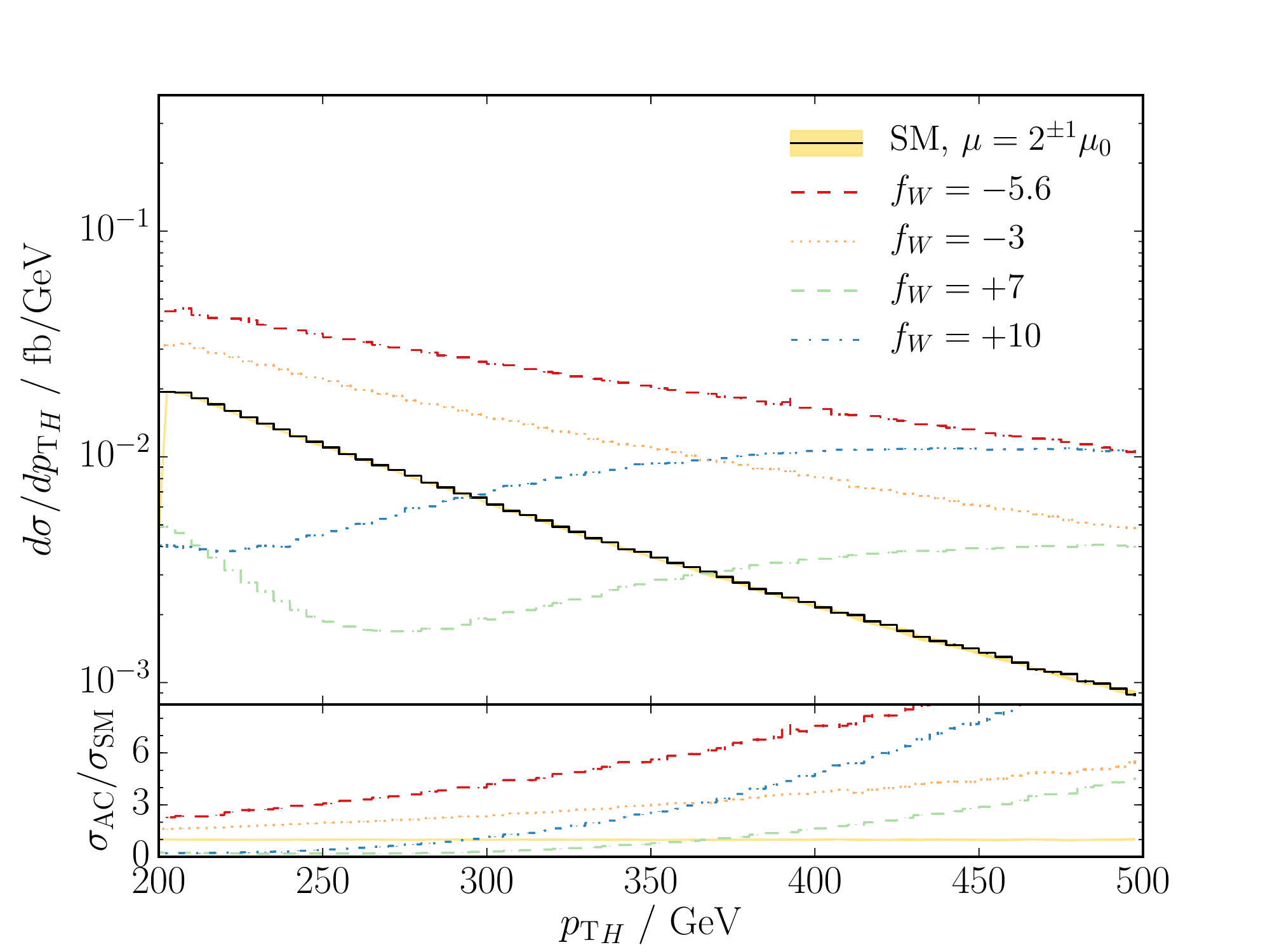}
		\caption{$\pt_H$, $W^+Hj$, boosted, with jet veto}
		\label{fig:whj-boost200-ac-ptz-nlo-lt02}
	\end{subfigure}
	\caption{Differential transverse momentum distribution of the $H$
          boson in \WH ($j$) production for different values of the $f_W$
          parameter with(right) and without(left) a dynamical jet veto.
					The AC scale is chosen as $\Lambda = \SI{1}{TeV}$.
}
	\label{fig:wh-boost200-ptw-ptz-nlo}
\end{figure*}

\begin{figure*}[htp]
	\centering
	\begin{subfigure}[htp]{\columnwidth}
		\centering
		\includegraphics[width=\columnwidth]{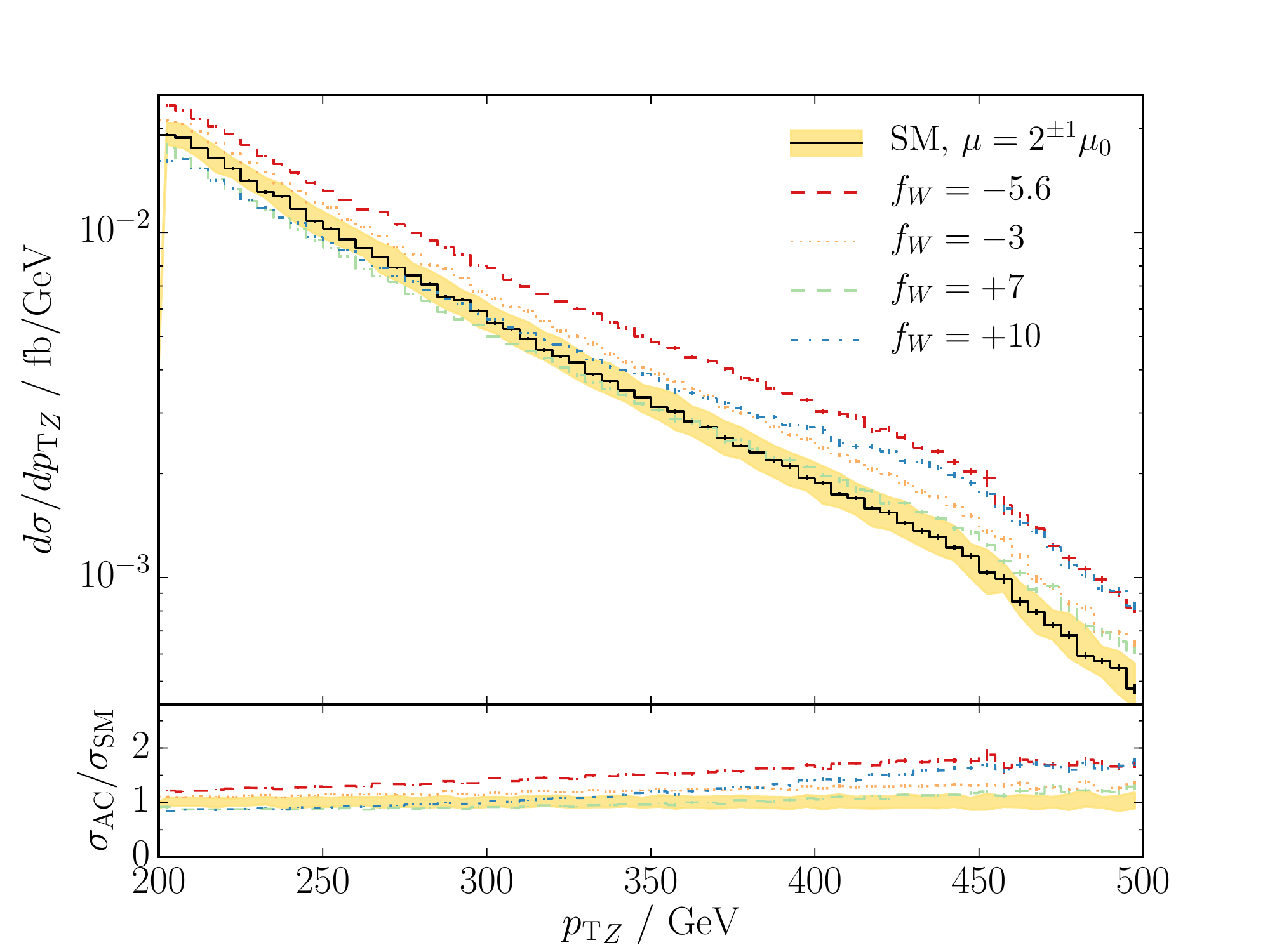}
		\caption{$\pt_Z$, $W^+Z$, boosted}
		\label{fig:wz-boost200-ac-ptz-nlo}
	\end{subfigure}
	\begin{subfigure}[htp]{\columnwidth}
		\centering
		\includegraphics[width=\columnwidth]{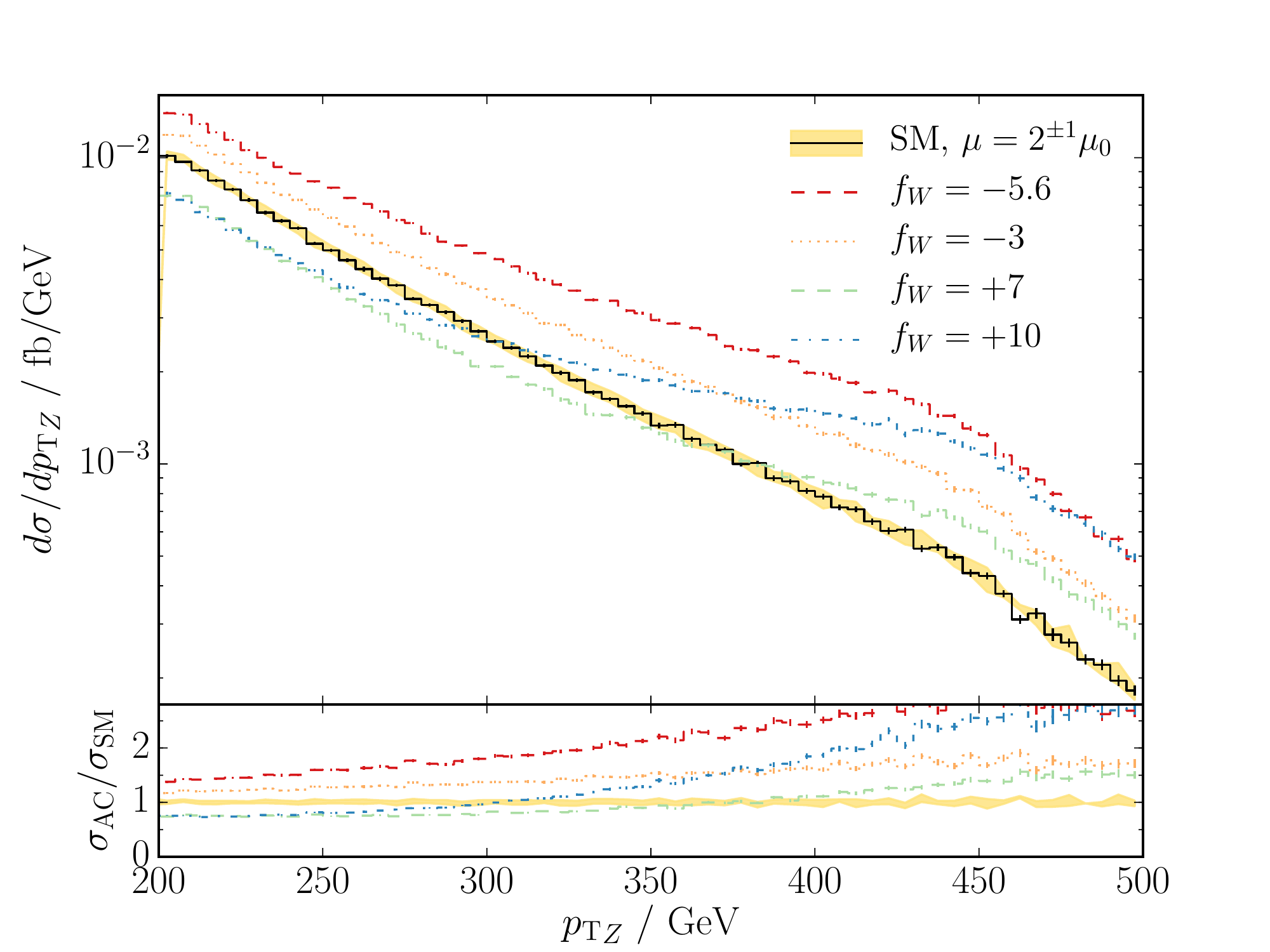}
		\caption{$\pt_Z$, $W^+Z$, boosted, with jet veto}
		\label{fig:wz-boost200-ac-ptz-nlo-lt02}
	\end{subfigure}
	\centering
	\begin{subfigure}[htp]{\columnwidth}
		\centering
		\includegraphics[width=\columnwidth]{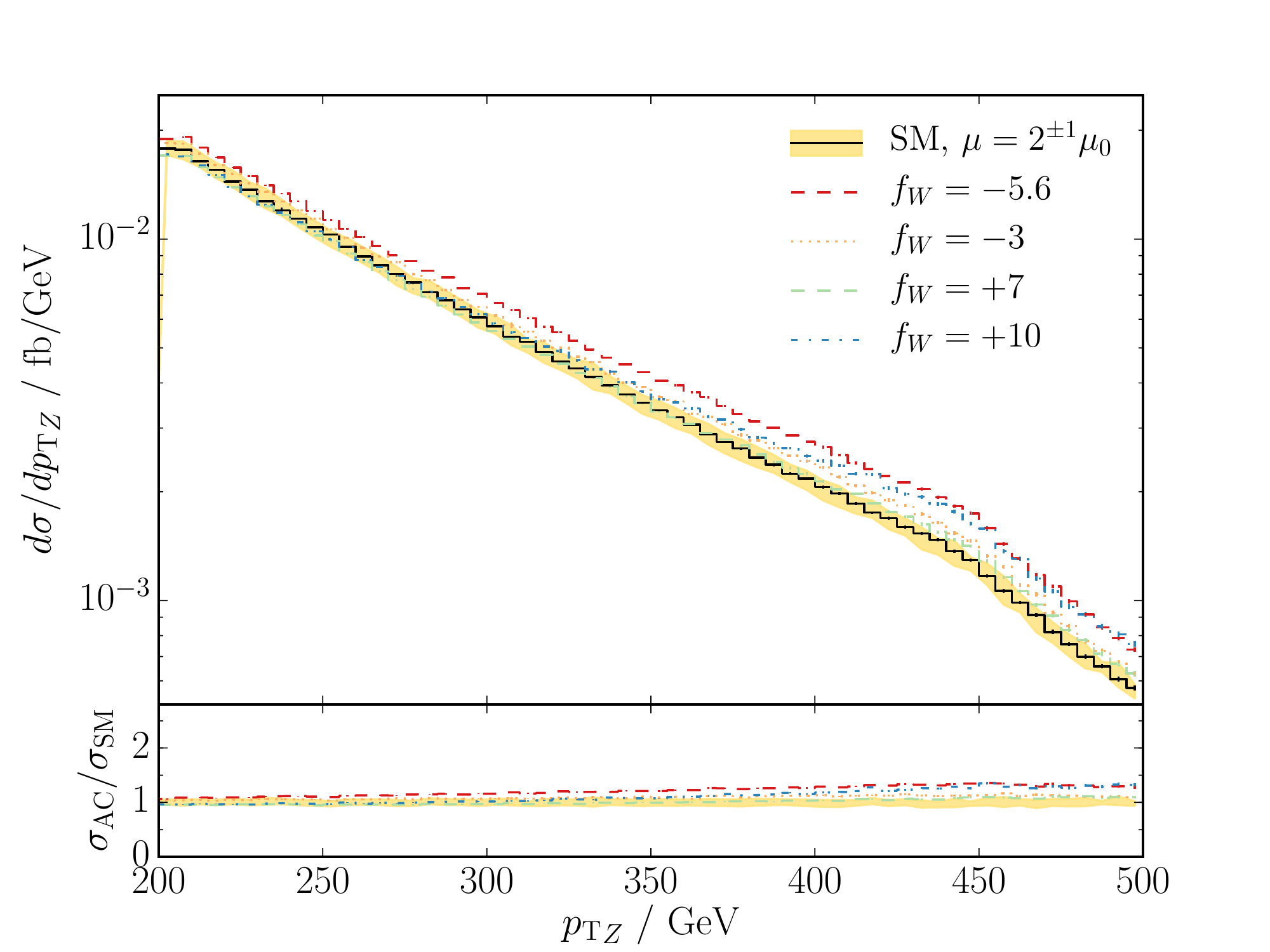}
		\caption{$\pt_Z$, $W^+Zj$, boosted}
		\label{fig:wzj-boost200-ac-ptz-nlo}
	\end{subfigure}
	\begin{subfigure}[htp]{\columnwidth}
		\centering
		\includegraphics[width=\columnwidth]{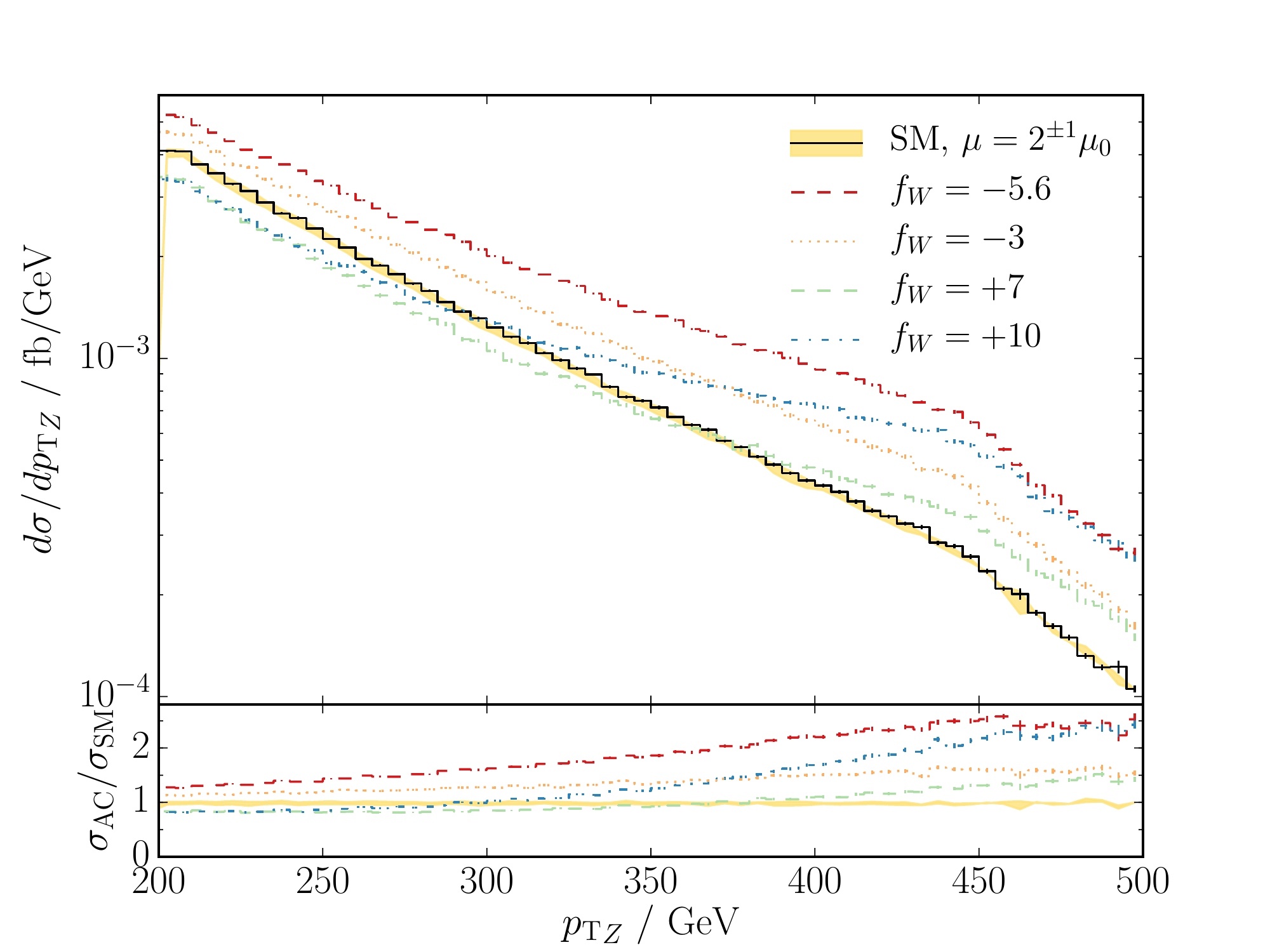}
		\caption{$\pt_Z$, $W^+Zj$, boosted, with jet veto}
		\label{fig:wzj-boost200-ac-ptz-nlo-lt02}
	\end{subfigure}
	\centering
	\begin{subfigure}[htp]{\columnwidth}
		\centering
		\includegraphics[width=\columnwidth]{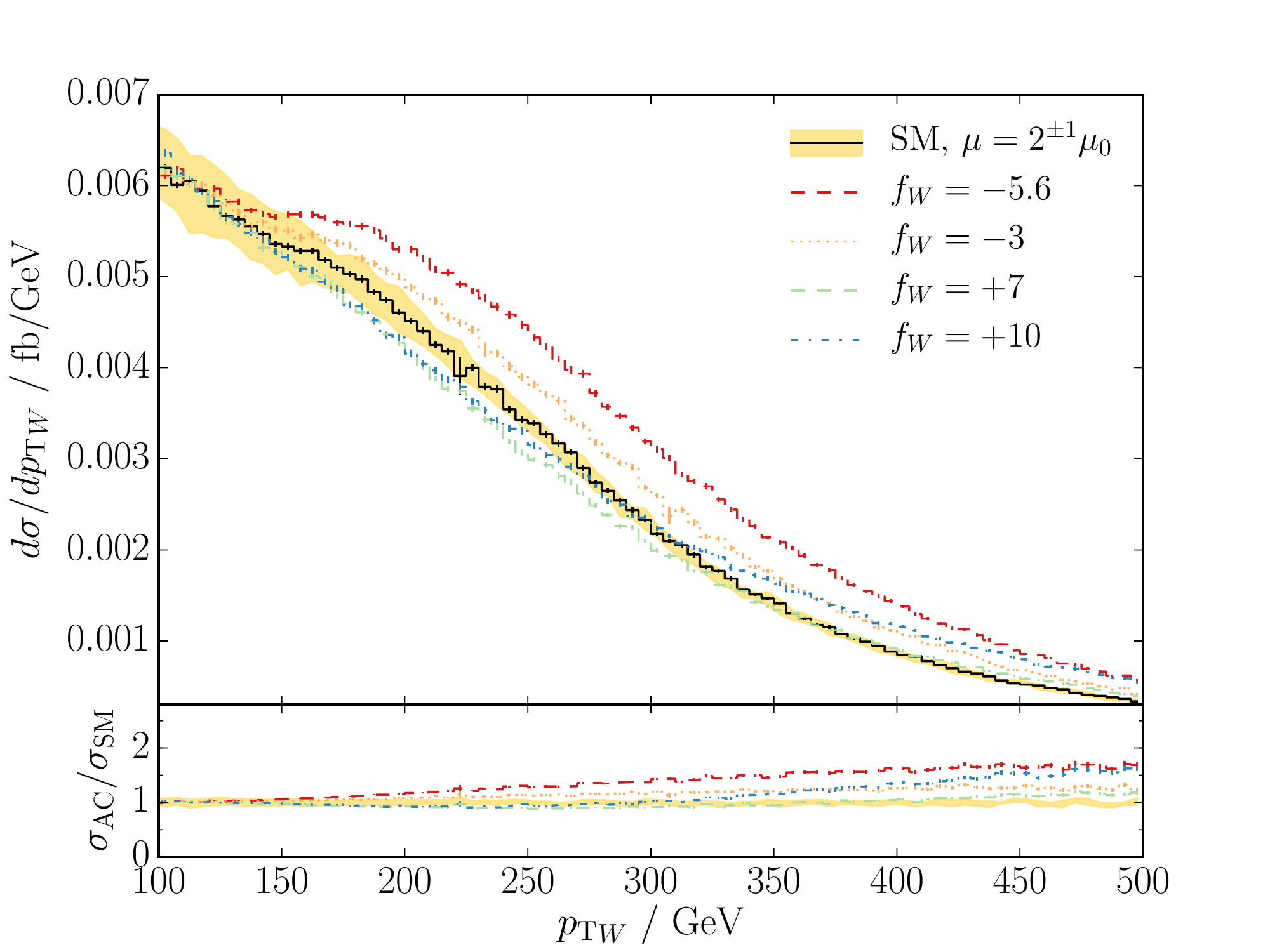}
		\caption{$\pt_W$, $W^+Zj$, boosted}
		\label{fig:wzj-boost200-ptw-nlo}
	\end{subfigure}
	\begin{subfigure}[htp]{\columnwidth}
		\centering
		\includegraphics[width=\columnwidth]{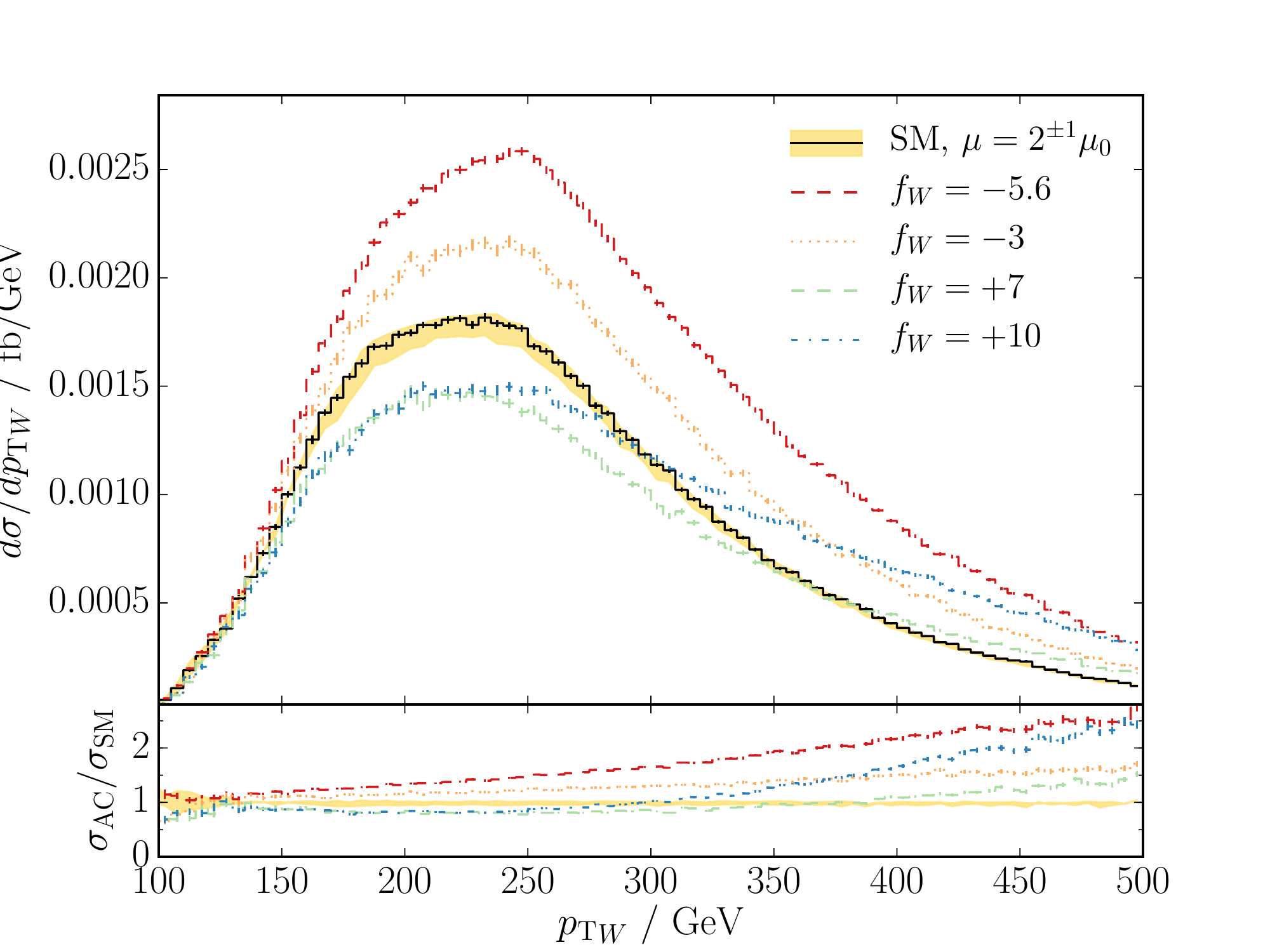}
		\caption{$\pt_W$, $W^+Zj$, boosted, with jet veto}
		\label{fig:wzj-boost200-ptw-nlo-lt02}
	\end{subfigure}

	\caption{Differential transverse momentum distribution of the
          reconstructed  $W$ and $Z$ boson in \WPZ ($j$) production for different values of the $f_W$
          parameter with(right) and without(left) a dynamical jet veto.
					The AC scale is chosen as $\Lambda = \SI{1}{TeV}$.
}
	\label{fig:wzj-boost200-ptw-ptz-nlo}
\end{figure*}

Fig.~\ref{fig:xjNLO} shows the same distributions at NLO\@. There is an
overall small shift to higher $x_{\text{jet}}$ due to the presence of an
additional parton in the real emission contributions. Still the jet dominated
and EW dominated phase space regions can be clearly separated.

Note that the $x_{\text{jet}} = 0.5$ and $x_Z = 0.5$ borders show 
unphysical structures at NLO due to phase space restrictions of the
1-parton final states, which affect the Born and virtual corrections.
In a final state with only one parton and thus exactly one massless jet 
after cuts, only $x_{\text{jet}} < 0.5$ is possible.
With two or more partons, the $x_{\text{jet}}$ definition allows values above
0.5, e.g.\ for two jets back-to-back with rather soft EW bosons or when there
is a parton not clustered into the jets.
Because this region is only available to the real emission at NLO and not to
the subtraction terms or the virtual corrections, there is an unphysical 
negative dip just below 0.5.
This problem affects an $x_{\text{jet}}$ definition based on transverse
momenta quite strongly,
while the definition using $\text{E}_{\text{T}}$ is safer since the 
masses act as a regulator.
In Fig.~\ref{fig:xjNLO}, the dip is barely visible.
A parton shower would completely wash out this artefact of the fixed order 
calculation.

Well below $x_{\text{jet}} = 0.5$, these infrared issues are mitigated. In
particular, the problem does not affect the region $x_{\text{jet}} < 0.2$, 
which is the region most sensitive to AC, as is visible in
Fig.~\ref{fig:xjNLOAC}, where we have used $f_W/\Lambda^2=\SI{-10}{TeV^{-2}}$
as an example. Note in the left panel of 
Fig.~\ref{fig:xjNLOAC} that the relative importance of hard jet events,
characteristic of the $WZj$ process, has diminished considerably once AC are
turned on, highlighting the fact that AC effects are more prominent in
back-to-back $WZ$ topologies.

Fig.~\ref{fig:whj-wzj-xj-ac} shows the 1D projection of the differential
$x_{\text{jet}}$ distribution for different values of the AC\@. For small
values of $f_W/\Lambda^2$,
the dominant term is the interference between SM and AC contribution.
Hence, their relative sign is important. 
For negative couplings there is constructive interference, while for positive
values of the coupling the interference is destructive. 
Thus, first the cross section decreases until the pure AC term outweighs the
interference term, which happens for $W^+Zj$ production 
at $f_W/\Lambda^2 \approx \SI[retain-explicit-plus]{+10}{TeV^{-2}}$.
Both Figs.~\ref{fig:xjNLOAC} and~\ref{fig:whj-wzj-xj-ac} show that the sensitivity to
AC effects is in the low $x_{\text{jet}}$ region, confirming that hard radiation dilutes the
sensitivity to AC searches. %
%
We will impose a jet veto requiring $x_{\text{jet}} < 0.2$ to focus on the region most
sensitive to AC\@.
There is small sensitivity to AC up to about $0.3$, such that this part of phase space
should also be included in experimental searches.

A fixed scale jet veto typically introduces logarithms of the veto scale 
over the hard process scale.
They are visible in form of a widening scale variation band for example in 
invariant mass or transverse momentum distributions as the energy increases.
An indication, that the dynamical jet veto does results in a more reliable
theoretical predictions than a veto above a fixed jet transverse momentum, 
is that the scale variation bands of vetoed cross sections do not grow with 
energy but stay at a size comparable to the total boosted sample. Such 
scale variation bands are shown in 
Figs.~\ref{fig:whj-wzj-xj-ac}--\ref{fig:wzj-boost200-ptw-ptz-nlo}.

\subsubsection*{Differential Distributions}

In Figs.~\ref{fig:wh-boost200-ptw-ptz-nlo}~(\ref{fig:wzj-boost200-ptw-ptz-nlo}), we show
transverse momentum differential distributions at NLO for $W^+H(j)$~($W^+Z(j)$) production.
$WX$ production at NLO includes 0-jet and 1-jet events, while $WXj$ production at NLO requires at
least one jet with $\pt_j>\SI{30}{GeV}$.

In Fig.~\ref{fig:wh-boost200-ptw-ptz-nlo}, we show the NLO differential 
distribution of the transverse momentum of the 
Higgs boson for the boosted sample with (right) and without (left) applying
the dynamical jet veto for $W^+H$ production (upper panels)
and for $W^+Hj$ production (lower panels).
The effect of AC is clearly visible in all distributions.
As expected, there is only a mild improvement when applying the
jet veto since there is little hard jet radiation in this process.

This situation needs to be contrasted with the case of $WZ$ production, as 
shown in Fig.~\ref{fig:wzj-boost200-ptw-ptz-nlo}. In the upper panel, we
consider the transverse momentum distribution of the 
$Z$ boson for $W^+Z$ production at NLO\@. 
Distributions are shown for the boosted sample with (right) and without (left) 
applying the dynamical jet veto. The effect of anomalous couplings is strongly 
enhanced by the jet veto.
Similarly, in the middle and lower rows, we show for $W^+Zj$ production,
the differential distribution of the transverse momenta of the $Z$ and
the $W$ bosons. Also here one can clearly see the improved AC sensitivity 
of the vetoed distributions. 
For $f_W = \SI{-5.6}{TeV^{-2}}$, in the case of $p_{TW}$, the ratio $\sigma_{\rm{AC}}/\sigma_{\rm{SM}}$ increases from 1.3 to
1.5 at \SI{250}{GeV} and from 1.7 to 2.8 at \SI{500}{GeV}.
For $p_{TZ}$, the increases at the same positions are from 1.1 to 1.5 and from 1.3 to 2.4.

\section{Conclusions}
\label{sec:con}
In this article, the QCD radiation patterns for the $WZ$ and the $WH$ production
processes have been studied. To accomplished this, we have computed and
implemented in \texttt{VBFNLO}, the $WH(j)$ production process at NLO QCD, including the
leptonic decay of the bosons as well as anomalous couplings effects. 

Looking at jet observables, we find distinguishable radiation patterns 
comparing $WH$ production with $WZ$ production.
While in $WH(j)$ production soft QCD radiation is preferred, in the $WZ(j)$ case with
boosted cuts there
are two equally important phase space regions, those with soft jets and those
with one hard vector boson recoiling against a jet and a second soft
vector boson. The latter region dilutes the sensitivity of this process to AC\@.
The two phase space regions can be separated quite cleanly by analyzing the
Dalitz-like normalized transverse energy fractions defined in
Eqs.~\ref{eq:xjdefet}~and~\ref{eq:xhzdefet}. 
 
To enhance the sensitivity to AC, a cut on the jet transverse energy fraction,
$x_{\text{jet}}$, proves effective. 
At the same time this dynamical jet veto provides more
reliable results than a fixed veto,
because it avoids large logarithms involving the veto scale 
and thus has smaller scale variations.

\begin{acknowledgments}
We would like to thank Johannes Bellm for helpful discussions and 
for the implementation of the $WZ$ production process in {\tt VBFNLO}.
We acknowledge the support from the Deutsche Forschungsgemeinschaft
via the Sonderforschungsbereich/Transregio SFB/TR-9 Computational
Particle Physics.  FC is funded by a Marie Curie fellowship
(PIEF-GA-2011-298960) and partially by MINECO (FPA2011-23596) and by
LHCPhenonet (PITN-GA-2010-264564).
RR is supported by the \emph{Landesgraduiertenf\"orderung des Landes Baden-W\"urttemberg}.
\end{acknowledgments}

\bibliographystyle{h-physrev} \bibliography{QCDVVjj}

\end{document}